\documentclass[twocolumn,preprintnumbers,showpacs,nofootinbib]{revtex4}

\usepackage{amssymb}
\usepackage{amsmath}
\usepackage{graphicx}

\newcommand{\sss}[1]{{\scriptscriptstyle{#1}}}
\newcommand{\boldmathsymbol}[1]{\ensuremath{\boldsymbol{#1}}}
\newcommand{\mean}[1]{\left\langle #1 \right\rangle}

\newcommand{\order}[1]{\mathcal{O}\!\left(#1\right)}

\newcommand{\Heaviside}[1]{\mathrm{H}\!\left(#1\right)}
\newcommand{\like}[2]{\mathcal{L}\!\left(#1 | #2 \right)}
\newcommand{\fiducial}[1]{\hat{#1}}

\newcommand{\CAMB}{\texttt{CAMB}}
\newcommand{\COSMOMC}{\texttt{CosmoMC}}

\newcommand{\cm}{\mathrm{cm}}
\newcommand{\km}{\mathrm{km}}
\newcommand{\meters}{\mathrm{m}}
\renewcommand{\sec}{\mathrm{s}}

\newcommand{\Mpc}{\mathrm{Mpc}}
\newcommand{\MHz}{\mathrm{MHz}}
\newcommand{\K}{\mathrm{K}}
\newcommand{\mK}{\mathrm{mK}}
\newcommand{\muK}{\mu\mathrm{K}}
\newcommand{\rad}{\mathrm{rad}}

\newcommand{\ucmb}{\mathrm{cmb}}
\newcommand{\urad}{\mathrm{rad}}
\newcommand{\ugas}{\mathrm{gas}}

\newcommand{\urec}{\mathrm{rec}}

\newcommand{\ureio}{\mathrm{reio}}
\newcommand{\uiono}{\mathrm{iono}}
\newcommand{\usys}{\mathrm{sys}}
\newcommand{\ucov}{\mathrm{cov}}
\newcommand{\uant}{\mathrm{ant}}
\newcommand{\ufov}{\mathrm{fov}}
\newcommand{\uexp}{\mathrm{exp}}
\newcommand{\usky}{\mathrm{sky}}
\newcommand{\ueff}{\mathrm{eff}}
\newcommand{\unoise}{\mathrm{noise}}
\newcommand{\uT}{\mathrm{T}}

\newcommand{\uB}{\mathrm{B}}
\newcommand{\uA}{\mathrm{A}}
\newcommand{\uH}{\mathrm{H}}
\newcommand{\uHI}{\mathrm{HI}}

\newcommand{\uHe}{\mathrm{He}}
\newcommand{\uda}{\mathrm{da}}
\newcommand{\uS}{\mathrm{S}}

\newcommand{\ub}{\mathrm{b}}
\newcommand{\un}{\mathrm{n}}
\newcommand{\up}{\mathrm{p}}
\newcommand{\udm}{\mathrm{dm}}

\newcommand{\uc}{\mathrm{c}}

\newcommand{\ui}{\mathrm{i}}
\newcommand{\us}{\mathrm{s}}
\newcommand{\dd}{\mathrm{d}}
\newcommand{\ud}{\dd}
\newcommand{\ee}{\mathrm{e}}
\newcommand{\ue}{\ee}

\newcommand{\calA}{\mathcal{A}}
\newcommand{\calH}{\mathcal{H}}
\newcommand{\calP}{\mathcal{P}}

\newcommand{\bk}{\boldmathsymbol{k}}
\newcommand{\bu}{\boldmathsymbol{u}}
\newcommand{\bv}{\boldmathsymbol{v}}
\newcommand{\bx}{\boldmathsymbol{x}}
\newcommand{\br}{\boldmathsymbol{r}}
\newcommand{\n}{\boldmathsymbol{\hat{n}}}

\newcommand{\GN}{G_{\sss{\mathrm N}}}
\newcommand{\Mp}{M_{\sss{\mathrm Pl}}}
\newcommand{\kB}{k_{\sss{\uB}}}

\newcommand{\nS}{n_{\sss{\uS}}}
\newcommand{\As}{A_{\sss{\uS}}}
\newcommand{\kstar}{k_*}
\newcommand{\calPs}{\calP_{\sss{\uS}}}

\newcommand{\OmegaR}{\Omega_\gamma}
\newcommand{\OmegaB}{\Omega_\ub}
\newcommand{\OmegaDM}{\Omega_\udm}

\newcommand{\Yp}{Y}
\newcommand{\fnHe}{f_{\sss{\uHe}}}

\newcommand{\Dang}{D_\uA}

\newcommand{\Esf}{E_{21}}
\newcommand{\nusf}{\nu_{21}}
\newcommand{\lambdasf}{\lambda_{21}}
\newcommand{\Asf}{A_{10}}
\newcommand{\tausf}{\tau_{21}}
\newcommand{\Tspin}{T_\us}
\newcommand{\Trad}{T_\urad}
\newcommand{\Tcmb}{T_\ucmb}
\newcommand{\Tgas}{T_\ugas}
\newcommand{\Tb}{T_\ub}
\newcommand{\Tn}{T_\un}
\newcommand{\Tbtilde}{\widetilde{T}_\ub}
\newcommand{\TE}{T_{\epsilon}}
\newcommand{\nH}{n_{\sss{\uHI}}} 
\newcommand{\nI}{n_\ui}          
\newcommand{\nb}{n_\ub}          
\newcommand{\nelec}{n_\ue}       

\newcommand{\xH}{x_{\sss{\uH}}}  
\newcommand{\xI}{x_{\sss{\ui}}}  
\newcommand{\xe}{x_{\ue}}        
\newcommand{\DeltaEoR}{\Delta_{\ureio}}
\newcommand{\zreio}{z_\ureio}
\newcommand{\DeltaHe}{\Delta_{\star}}
\newcommand{\zHe}{z_{\star}}
\newcommand{\Deltaz}{\Delta_z}
\newcommand{\fidDeltaz}{\fiducial{\Delta}_z}
\newcommand{\TspinMax}{\Tspin^{\max}}
\newcommand{\fidTspinMax}{\fiducial{T}_\us^{\max}}
\newcommand{\TspinRec}{\Tspin^{\uda}}

\newcommand{\mH}{m_{\sss{\uH}}}
\newcommand{\mHe}{m_{\sss{\uHe}}}
\newcommand{\E}{\epsilon}
\newcommand{\tauE}{\tau_{\epsilon}}
\newcommand{\rscatE}{r_\epsilon}
\newcommand{\etaE}{\eta_{\epsilon}}
\newcommand{\tauC}{\tau_\uc}
\newcommand{\sigmaT}{\sigma_{\sss{\uT}}}

\newcommand{\D}{\Delta}
\newcommand{\Dmono}{\D_\us}
\newcommand{\DHI}{\D_{\sss{\uHI}}}

\newcommand{\DTspin}{\D_{\Tspin}}
\newcommand{\DTrad}{\D_{\Trad}}
\newcommand{\DxH}{\D_{\xH}}
\newcommand{\DxI}{\D_{\xI}}
\newcommand{\Db}{\D_\ub}
\newcommand{\Dv}{\D_v}
\newcommand{\dTb}{\delta \Tb}

\newcommand{\Pk}{P}
\newcommand{\calPk}{\mathcal{\Pk}}
\newcommand{\PTb}{\Pk_{\dTb}}
\newcommand{\Pb}{\Pk_{\ub}}
\newcommand{\Pii}{\Pk_{\ui\ui}}
\newcommand{\Pbi}{\Pk_{\ui\ub}}
\newcommand{\Pib}{\Pbi}
\newcommand{\Pv}{\Pk_{v}}
\newcommand{\Pvi}{\Pk_{v\ui}}
\newcommand{\Piv}{\Pvi}
\newcommand{\Pvb}{\Pk_{v\ub}}
\newcommand{\Pbv}{\Pvb}

\newcommand{\Pzero}{\Pk_{0}}
\newcommand{\Ptwo}{\Pk_{2}}
\newcommand{\Pfour}{\Pk_{4}}

\newcommand{\calPTb}{\calPk_{\dTb}}
\newcommand{\calPzero}{\calPk_{0}}
\newcommand{\calPtwo}{\calPk_{2}}
\newcommand{\calPfour}{\calPk_{4}}

\newcommand{\Nii}{N_{\ui\ui}}
\newcommand{\alphaii}{\alpha_{\ui\ui}}
\newcommand{\Rii}{R_{\ui\ui}}
\newcommand{\gammaii}{\gamma_{\ui\ui}}

\newcommand{\Nib}{N_{\ui\ub}}
\newcommand{\alphaib}{\alpha_{\ui\ub}}
\newcommand{\Rib}{R_{\ui\ub}}

\newcommand{\skyang}{\Theta}
\newcommand{\bskyang}{\boldmathsymbol{\skyang}}
\newcommand{\Freq}{F}

\newcommand{\brbot}{\boldmathsymbol{r}_\perp}
\newcommand{\rpara}{r_\parallel}
\newcommand{\ubot}{u_\perp}
\newcommand{\bubot}{\boldmathsymbol{u}_\perp}
\newcommand{\ubotmin}{\ubot^{\min}}
\newcommand{\ubotmax}{\ubot^{\max}}
\newcommand{\upara}{u_\parallel}

\newcommand{\uparamax}{\upara^{\max}}
\newcommand{\kbot}{k_\perp}
\newcommand{\bkbot}{\boldmathsymbol{k}_\perp}
\newcommand{\kpara}{k_\parallel}

\newcommand{\fwhm}{\theta_{\mathrm{fw}}}
\newcommand{\lbeam}{\ell_\ub}
\newcommand{\Tsys}{T_\usys}
\newcommand{\bwidth}{B}

\newcommand{\texp}{t_\uexp}
\newcommand{\fcover}{f_\ucov}
\newcommand{\fsky}{f_\usky}
\newcommand{\OmegaFov}{\Omega_\ufov}
\newcommand{\OmegaSky}{\Omega_\usky}
\newcommand{\Na}{N_\uant}
\newcommand{\Snu}{S_\nu}
\newcommand{\Aone}{\calA_1}
\newcommand{\Aeff}{\calA_\ueff}

\newcommand{\Csys}{C_\usys}
\newcommand{\Cnoise}{C^\unoise}
\newcommand{\Pnoise}{\Pk^{\unoise}}

\newcommand{\Aiono}{A_\uiono}
\newcommand{\dOmega}{\delta \Omega}

\newcommand{\dTn}{\delta \Tn}
\newcommand{\dV}{\delta V}
\newcommand{\nup}{\nu_\up}

\newcommand{\fish}{F}
\newcommand{\bfish}{\boldmathsymbol{\fish}}
\newcommand{\Nc}{N_\uc}
\newcommand{\fidlambda}{\fiducial{\lambda}}
\newcommand{\fidC}{\fiducial{C}}
\newcommand{\fidD}{\fiducial{D}}

\begin{document}

\title{Background reionization history from omniscopes}

\author{S\'ebastien Clesse} \email{s.clesse@damtp.cam.ac.uk}
\affiliation{DAMTP, Centre for Mathematical Sciences, Wilberforce
  Road, Cambridge CB3 0WA, United Kingdom}

\author{Laura Lopez-Honorez} \email{laura.lopez-honorez@mpi-hd.mpg.de}
\affiliation{Max-Planck-Institute for Nuclear Physics, Saupfercheckweg
  1, 69117 Heidelberg, Germany}
\affiliation{Theoretische Natuurkunde, Vrije Universiteit Brussel and
The International Solvay Institutes, Pleinlaan 2, B-1050 Brussels, Belgium}

\author{Christophe Ringeval} \email{christophe.ringeval@uclouvain.be}
\affiliation{Centre for Cosmology, Particle Physics and Phenomenology,
  Institute of Mathematics and Physics, Louvain University, 2 Chemin
  du Cyclotron, 1348 Louvain-la-Neuve, Belgium}

\author{Hiroyuki Tashiro} \email{hiroyuki.tashiro@asu.edu}
\affiliation{Physics Department, Arizona State University, Tempe, AZ
  85287, USA}

\author{Michel~H.~G. Tytgat} \email{mtytgat@ulb.ac.be} \affiliation{Service
  de Physique Th\'eorique, CP225 Universit\'e Libre de Bruxelles,
  Boulevard du Triomphe, 1050 Brussels, Belgium}

\date{\today}

\begin{abstract}
  The measurements of the 21-cm brightness temperature fluctuations
  from the neutral hydrogen at the Epoch of Reionization (EoR) should
  inaugurate the next generation of cosmological observables. In this
  respect, many works have concentrated on the disambiguation of the
  cosmological signals from the dominant reionization
  foregrounds. However, even after perfect foregrounds removal, our
  ignorance on the background reionization history can significantly
  affect the cosmological parameter estimation. In particular, the
  interdependence between the hydrogen ionized fraction, the baryon
  density and the optical depth to the redshift of observation induce
  nontrivial degeneracies between the cosmological parameters that
  have not been considered so far. Using a simple, but consistent
  reionization model, we revisit their expected constraints for a
  futuristic giant 21-cm omniscope by using for the first time
  Markov Chain Monte Carlo (MCMC) methods on multiredshift full
  sky simulated data. Our results agree well with the usual Fisher
  matrix analysis on the three-dimensional flat sky power spectrum but
  only when the above-mentioned degeneracies are kept under
  control. In the opposite situation, Fisher results can be
  inaccurate. We show that these conditions can be fulfilled by
  combining cosmic microwave background measurements with multiple
  observation redshifts probing the beginning of EoR. This allows a
  precise reconstruction of the total optical depth, reionization
  duration and maximal spin temperature. Finally, we discuss the
  robustness of these results in presence of unresolved ionizing
  sources. Although most of the standard cosmological parameters
  remain weakly affected, we find a significant degradation of the
  background reionization parameter estimation in presence of nuisance
  ionizing sources.
\end{abstract}


\maketitle

\section{Introduction}

\label{sec:intro}

Among the next generation of cosmological probes, interferometric
radio telescopes observing the redshifted 21-cm line associated with
the hyperfine transitions of neutral hydrogen atoms have attracted a
lot of attention (see Refs.~\cite{Loeb:2000fc, Furlanetto:2006jb,
  Barkana:2006ep, Pritchard:2011xb} for reviews). These telescopes are
revolutionary in their design as in the proposed Fast Fourier
Transform Telescope (FFTT) which is conceptually an all digital
antennas array imaging the whole visible sky at
once~\cite{Tegmark:2008au}. Images would then be reconstructed by a
two-dimensional fast Fourier transform over the $N$ antennas signal,
i.e. in $N \ln N$ operations. As argued in
Ref.~\cite{Tegmark:2009kv}, compared to the required $N^2$ pairing in
traditional interferometers, the gain could be used to scale up the
telescope size and sensitivity thereby allowing measurements of
cosmological signals. The first generation of such telescopes is under
deployment~\cite{Peterson:2006bx, Garrett:2009gp,
  Lazio:2009bea,Rawlings:2011dd, Pen:2008ut, Mitchell:2010cc,
  Ord:2010st}. Although not designed for cosmological purposes, they
aim at detecting a cosmological signal from the Epoch of
reionization~\cite{Parsons:2009in, Bittner:2010fi, Santos:2010hj,
  Harker:2010ht, Chang:2010jp, Paciga:2010yy, Morandi:2011hn}. It is
still a matter of active research to know if the signal coming from
the astrophysical foregrounds can be properly separated from the
cosmological one~\cite{DiMatteo:2001gg, Jelic:2008jg, Bowman:2008mk,
  Petrovic:2010me, Harker:2011et}. One should indeed keep in mind that
the former is actually a few order of magnitude stronger than the
latter and this has triggered interest in the cross-correlation of
the 21-cm signal with much cleaner data such as the Cosmic Microwave
Background (CMB) or galaxy surveys~\cite{Alvarez:2005sa,
  Tashiro:2009uj, Jelic:2009sd,Wyithe:2006vg, Furlanetto:2006pg}.

\begin{figure}
  \begin{center}
    \includegraphics[width=\columnwidth]{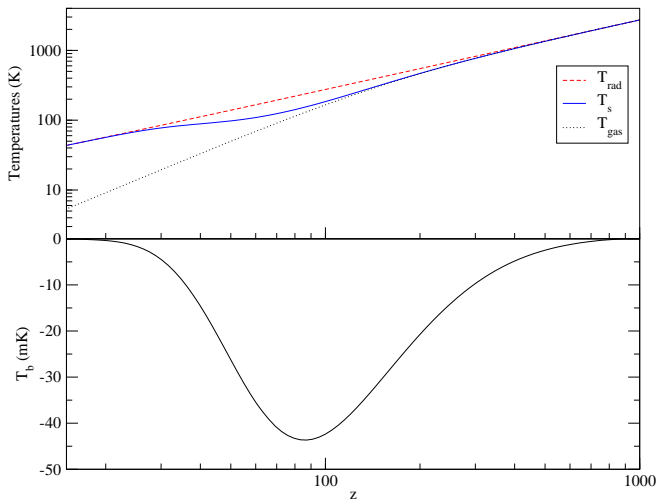}
    \caption{Background evolution of the neutral hydrogen gas and spin
      temperatures (top) as a function of the redshift during the dark
      ages. The spin temperature $\Tspin$ being defined as $n_1/n_0 =
      3 \exp[-\hbar \omega_{21}/(k_B\Tspin)]$ where $n_0$ and $n_1$
      denote the number density of atoms in the singlet and triplet
      hyperfine states, respectively. The bottom panel shows the
      resulting brightness temperature.}
    \label{fig:btbg}
  \end{center}
\end{figure}

The physical origin of the cosmological 21-cm radiation lies in the
differential cooling induced by the expansion of the Universe on
relativistic and non-relativistic gases. After recombination, one
would naively expect  the temperature of neutral hydrogen $\Tgas$ to
decrease in $1/a^2$, where $a$ is the
Friedmann--Lema\^{\i}tre--Robertson--Walker (FLRW) scale factor, which
is faster than the radiation temperature $\Trad$ scaling as $1/a$. As
a result, the neutral hydrogen gas becomes cool enough to be able to
absorb CMB photons by a spin flip hyperfine transition at a wavelength
of 21-cm. Tuning a radio telescope at the corresponding redshifted
frequency allows, in principle, to probe the density fluctuations of
neutral hydrogen over the sky and at any redshift, hence the so-called
cosmological tomography~\cite{1979MNRAS.188..791H,
  1990MNRAS.247..510S}. In reality, the situation is a bit more
complex and one has to take into account the evolution of the
Boltzmann distribution of neutral hydrogen due to the collisions with
the residual electrons, protons and absorption versus stimulated
emission of CMB photons~\cite{Madau:1996cs, Seager:1999bc,
  Seager:1999km, Furlanetto:2007te}. The background evolution can
nevertheless be numerically derived and we have plotted in
Fig.~\ref{fig:btbg} the resulting brightness temperature evolution.

For an assumed generic set of cosmological parameters, taken from the
Wilkinson Microwave Anisotropy Probe (WMAP) seven-year
data~\cite{Komatsu:2010fb}, the absorption is maximal around a
redshift of $z\simeq 10^2$. Notice that at lower redshifts, the spin
temperature is driven again towards the photon temperature and the
signal vanishes till the EoR. The evolution of neutral hydrogen
density fluctuations can similarly been predicted during the dark ages
by using the theory of cosmological perturbations~\cite{Loeb:2003ya,
  Bharadwaj:2004nr, Naoz:2005pd}. Provided the length scales of
interest remain in the linear regime, the theoretical predictions for
the 21-cm power spectra are neat~\cite{Hirata:2006bn, Lewis:2007kz,
  Lewis:2007zh, Mao:2008ug}. Being three-dimensional in nature, the
information content is huge and have been used to forecast constraints
on various cosmological models such as
non-Gaussianities~\cite{Pillepich:2006fj, Cooray:2006km,
  Joudaki:2011sv,Chongchitnan:2012we}, cosmic
strings~\cite{Khatri:2008zw, Brandenberger:2010hn, Berndsen:2010xc,
  Pagano:2012cx}, dark matter signatures~\cite{Shchekinov:2006eb,
  Valdes:2007cu, Borriello:2008gy, Cumberbatch:2008rh,
  Natarajan:2009bm}, modified gravity~\cite{Brax:2012cr} and
inflation~\cite{Barger:2008ii, Gordon:2009wx, Masui:2010cz,
  Adshead:2010mc}.

\begin{figure}
  \begin{center}
    \includegraphics[width=\columnwidth]{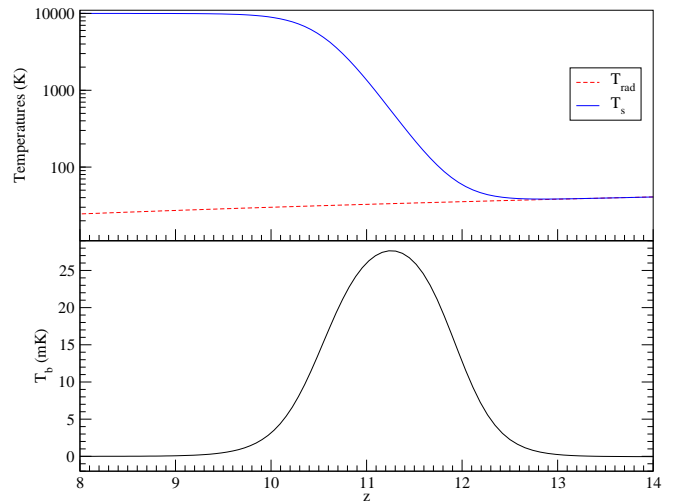}
    \caption{Reionization driven spin temperature (top) and brightness
      temperature (bottom) with respect to the observation redshift.
      The reionization model is described in Sec.~\ref{sec:eormodel}
      ($\tau = 0.088$).}
    \label{fig:btbgeor}
  \end{center}
\end{figure}

As already mentioned, in addition to having a small amplitude compared
to foregrounds, the dark ages signal is redshifted to wavelengths of
typically twenty meters and it is not obvious that it may actually be
used for cosmology in a foreseeable future. One very futuristic
approach would be to build those radio telescopes on the
Moon~\cite{Jester:2009dw}. A more reasonable approach is the
possibility to probe the hydrogen density fluctuations with the 21-cm
line at the EoR~\cite{Ciardi:2003hg, Sethi:2005gv,
  Barkana:2005xu}. This time, the exciting photons are coming from
reionization sources (stars or quasars) instead of the CMB and are
expected to produce an efficient spin states population
inversion. This results into $\Tspin \gg \Trad$ and generates an
emission line at a much smaller redshifted wavelength. We have plotted
the expected evolution of the brightness temperature during the EoR in
Fig.~\ref{fig:btbgeor} for a simple reionization model that we
describe in Sec.~\ref{sec:eormodel}.

The resulting signal is around $25\,\mK$ in amplitude, which is of the
same order than those coming from the dark ages, but at a higher
frequency thereby rendering its detection more likely. However, the
theoretical predictions are now far less neat due to the additional
astrophysical uncertainties associated with the way reionization
proceeds~\cite{2011MNRAS.410.1377T, Mao:2011xp}. However since
$21\,\cm$ data are three-dimensional by nature, it has been shown in
Refs.~\cite{Liu:2011hh, Liu:2011ih} that the redshift evolution could
be used to efficiently eliminate the expected foregrounds while still
keeping some of the so-called longitudinal modes for cosmology.

Forecasting the 21-cm constraints for cosmology is usually made from
Fisher matrix analysis which merely assumes the likelihood to be
Gaussian around the best fit. Provided the parameters are well
constrained, the method is fast and
accurate~\cite{Tegmark:1996bz}. However, as noted in
Ref.~\cite{Mao:2008ug}, some model parameters linked to reionization,
such as the optical depth or the hydrogen ionized fraction, can be
completely degenerated from the 21-cm point of view and factorize out
of the Fisher matrix. In a realistic situation, this should not be the
case as all reionization parameters are uniquely determined by the
background reionization history. One may therefore wonder how their
correlations to the standard cosmological parameters affect the
forecasts.

In Refs.~\cite{Pritchard:2010pa, Morandi:2011hn}, the problem of
reconstructing the background reionization history is specifically
addressed in the context of future 21-cm experiments that would be
devoted to the global signal, i.e. the homogeneous mode. Here, we
would like to discuss this issue for the 21-cm FFTT-like experiments
which are poorly, if not at all, sensitive to the constant mode. In that
situation, the cosmological brightness fluctuations $\dTb(\bx,z)$ have
simultaneously a dependence in both the background evolution and
spectral shape. This question is peculiar to 21-cm observables because
the homogeneous mode $\Tb(z)$ depends on various cosmological
parameters, and when concerned with EoR, on the reionization (see
Sec.\ref{sec:eorsignal}). Let us notice that such a situation does not
occur for CMB not only because $\Tcmb$ is measured, but because it is
uniquely determined by $\OmegaR$.

In this context, we would like to quantify how cosmological forecasts
are affected by the background reionization history, and if it is
possible for omniscopes to constrain the reionization parameters. For
this purpose, we consider a giant FFTT-like ground-based instrument
and assume that most of the foregrounds can indeed be eliminated using
redshift evolution. As a result, we will be keeping only a few
redshift slices for our forecasts, eventually marginalizing over some
nuisance ionization spectra modelled as in Ref.~\cite{Mao:2008ug}. We
consider a simple reionization model for which the time evolution of
the spin temperature, hydrogen ionized fraction and optical depth are
given as a function of some extra parameters. In this manner, it is
imposed that they are intrinsically correlated by reionization
physics. As already mentioned, in order to go beyond Fisher matrix, we
use MCMC methods over a modified version of the $\CAMB$ 21-cm
code~\cite{Lewis:1999bs, Lewis:2007kz, Lewis:2007zh} which includes
the Epoch of Reionization as well as some nuisance ionization
spectra. This allows us to use and compare both Fisher matrix analysis
and full sky simulated data coupled to MCMC
methods~\cite{Lewis:2002ah}.

The paper is organized as follows. In Sec.~\ref{sec:models}, we
present the reionization and telescope models used in the analysis
while the methodology and results are presented in
Sec.~\ref{sec:reiocast}. Starting from one redshift slice only, we
show that the reionization induced degeneracies damage, and sometime
prevent, the determination of some cosmological parameters, such as
the baryon density. In these situations, we also find that a Fisher
analysis can be inaccurate for some strongly correlated
parameters. Combining a few redshifts, which sample the beginning of
reionization, improves the situation but does not compete with a
Planck-like CMB experiment. However, we show in Sec.~\ref{sec:addcmb},
that combining 21-cm and CMB lifts almost all those degeneracies
thereby improving the overall accuracy by an order of magnitude. In
that situation our MCMC results match with the usual Fisher matrix
analysis while the underlying reionization model can be completely
determined. Finally, we show that these results may be mitigated by
the presence of toy nuisance reionization spectra.

\section{Reionization and telescope models}
\label{sec:models}

\subsection{21-cm signal from the EoR}
\label{sec:eorsignal}

\subsubsection{Background}

Assuming a small emission line profile, the 21-cm brightness
temperature during the EoR and at a given
observed energy $\E$ is given by~\cite{Lewis:2007kz}
\begin{equation} 
\label{eq:Tb}
  \Tb(\E) = \left( 1 -  \ue ^{- \tauE} \right)
  \left. \frac{\Tspin - \Trad }{1+z} \right|_{\etaE} \, ,
\end{equation}
where  $\tauE$ is the optical depth to 21-cm
\begin{equation}
\label{eq:tauE}
  \tauE = \left.\frac{3 \hbar c^3 \Asf  \nH(\eta)} {16 \kB \nusf^2
    \Tspin(\eta)  H(\eta)}\right|_{\etaE}.
\end{equation}
All quantities are evaluated at the conformal time $\eta = \etaE$ at
which $\epsilon = a(\etaE) \Esf=2\pi a(\etaE) \hbar \nusf$, $\nusf$
being the rest frame frequency of the 21-cm spin flip atomic
transition and $a$ is the scale factor. In the following, we use
natural units $\kB=\hbar=c=1$. In Eq.~(\ref{eq:tauE}), $\Asf \simeq
2.869 \times 10^{-15}\,\us^{-1} $ is the Einstein coefficient of
spontaneous emission, $\nH(\eta)$ the density of neutral hydrogen
atoms and $H(\eta)$ the Hubble parameter. Introducing the neutral
fraction $\xH \equiv \nH/(\nH + \nI)$, with $\nI$ the number density
of ionized hydrogen, one can express the density of neutral hydrogen
in terms of the cosmological parameters today as
\begin{equation}
\nH(z) = \xH(z) \dfrac{3 \OmegaB H_0^2 (1-\Yp)}{ 8 \pi \GN \mH}
\left(1+z\right)^3\,,
\end{equation}
where $\Yp\simeq 0.24$ is the helium mass fraction\footnote{Notice
  that the helium fraction affects significantly the brightness
  temperature for a given $\OmegaB$.} and $\mH$ the mass of the
hydrogen atom. During reionization, one assumes that $\Tspin$ is
driven to high values due to ionizing sources~\cite{Mao:2011xp} such
that one can expand Eq.~(\ref{eq:Tb}) in
\begin{equation}
\label{eq:Tbexpand}
\Tb = a \TE - a \dfrac{\TE}{\Tspin} \left(\Trad + \dfrac{1}{2} \TE
\right) + \order{\dfrac{\TE}{\Tspin}}^2,
\end{equation}
where 
\begin{equation}
\label{eq:TE}
\TE \equiv \tauE \Tspin = \dfrac{9 \Asf \Mp^2}{16 \nusf^2 \mH}
\dfrac{\OmegaB H_0^2 (1-\Yp)}{a^3(\etaE) H(\etaE)}\, \xH(\etaE) \,.
\end{equation}
In the limit $\Tspin \gg \TE$, $\Tb \simeq a \TE$. The brightness
temperature is thus positive (emission line), does not depend anymore
on the spin temperature and is directly proportional to $\xH$. Up to
the metric fluctuations, this is also the case for the linear
perturbations $\delta \Tb$ which directly probe the neutral hydrogen
fluctuations. However, the approximation $\Tspin \gg \TE$ cannot hold
at the beginning of reionization as $\Tspin$ should continuously rise
from the dark ages value to its maximum value deep in the reionization
era. As a result, it is far from evident that there exists a redshift
range for which this approximation is fully consistent. Actually, even
for a spin temperature $\Tspin \sim 10^3\,\K$, one can check that
differences at the percent level already arise between
Eq.~(\ref{eq:Tb}) and the leading term of Eq.~(\ref{eq:Tbexpand}). For
these reasons, in this paper, we keep the complete dependence in
$\Tspin$ and $\Trad$ and assume a simple, but consistent, model of
reionization to parametrize the spin temperature evolution (see
Fig.~\ref{fig:btbgeor} and Sec.~\ref{sec:eormodel}).

\subsubsection{Linear perturbations}

The brightness temperature fluctuations can be obtained by solving the
Boltzmann equations driving the evolution of the 21-cm photons in a
perturbed FLRW space-time. This has already been done by Lewis and
Challinor during the dark ages in Ref.~\cite{Lewis:2007kz}. We have
followed their approach while adding the contributions of the hydrogen
ionized fraction perturbations coming from the reionization
sources. For this purpose, we have modified the publicly available
$\CAMB$ code~\cite{Lewis:1999bs, Challinor:2011bk} and numerically
solved the full Boltzmann equations incorporating all the linear order
effects discussed in Ref.~\cite{Lewis:2007kz}, i.e. neglecting only
anisotropies higher than dipole photon ones and any broadening of the
emission line profile. The multipoles components of the 21-cm
distribution function actually used in the next sections have been
written down in Appendix~\ref{sec:boltz}.

In order to illustrate the physical processes at work, it is
nevertheless convenient to approximate the perturbed Boltzmann
distribution in the small scale limit and at leading order in
$\tauE$. Let us emphasize again that these approximations are not made
in the actual forecasts of Sec.~\ref{sec:reiocast}. Along the line-of-sight direction $\n$, at the measured energy $\E$, one gets
\begin{equation}
\label{eq:deltaTb}
\delta \Tb(\bx,\n,\E) \simeq \Tb \ue^{-\tauC} \left[ \Dmono - \dfrac{1}{a H}
\n\cdot \dfrac{\partial \bv}{\partial \eta}  \right]_{\etaE},
\end{equation}
where the $\Dmono$ is monopole source
\begin{equation}
\label{eq:Dmono}
\Dmono \equiv \DHI + \dfrac{\Trad}{\Tspin-\Trad} \left(\DTspin -
\DTrad \right),
\end{equation}
and all $\D_x \equiv \delta x/x$ stand for relative perturbations. The
second term of Eq.~(\ref{eq:deltaTb}) encodes the redshift distortions
due to the perturbations $\bv$ in the (baryonic) gas relative velocity
to the observer. As shown in Ref.~\cite{Lewis:2007kz}, Thompson
scattering suppresses the 21-cm brightness on all scales and is
responsible of the exponential term in Eq.~(\ref{eq:deltaTb}),
$\tauC(\etaE)$ being the Thompson optical depth to the redshift of
observation (not to be confused with $\tauE$). During the dark ages,
$\xH \simeq 1$ and the perturbations $\DHI$ in the neutral hydrogen
directly trace the baryonic perturbations $\Db\equiv \delta
\nb/\nb$. At the EoR, the ionizing sources on their own are expected
to induce new perturbations in the neutral fraction such that $\DHI =
\DxH + \Db$ (at fixed helium fraction), or, in terms of the
\emph{hydrogen} ionized fraction $\xI$ (not to be confused with the
total ionized fraction $\xe$)
\begin{equation}
\label{eq:DHI}
\DHI = \Db - \dfrac{\xI}{\xH} \DxI\,,
\end{equation}
where $\DxI$ is the relative ionized fraction perturbation. Notice
that the ionizing sources are also generating additional spin
temperature perturbations through their effects on the gas temperature
and Lyman-$\alpha$
emission~\cite{Pritchard:2008da,Santos:2010hj,2011MNRAS.410.1377T}. However,
they are not considered here for simplicity. The expression for the
brightness fluctuations can be further simplified by assuming that the
regime $\Tspin \gg \Trad$ holds at the redshift of interest. In this
limit, Eq.~(\ref{eq:deltaTb}) reads in Fourier space
\begin{equation}
\label{eq:deltaTbhot}
\delta \Tb(\bk,\n,\E) \simeq \ue^{-\tauC}  \Tbtilde(\E) \left[\xH \Db - \xI \DxI -
  \mu^2 \xH \Dv \right]_{\etaE},
\end{equation}
where $\mu \equiv \bk \cdot \n/k$ and $\Tbtilde \equiv \Tb/\xH$ is the
background brightness temperature if hydrogen were fully neutral,
i.e. from Eq.~(\ref{eq:Tbexpand}) and (\ref{eq:TE})
\begin{equation}
\label{eq:Tbtilde}
\Tbtilde(\E) = \dfrac{9 \Asf \Mp^2}{16 \nusf^2 \mH}
\dfrac{\OmegaB H_0^2 (1-\Yp)}{a^2(\etaE) H(\etaE)}\,.
\end{equation}
In Eq.~(\ref{eq:deltaTbhot}), we have introduced the perturbed
quantity
\begin{equation}
\label{eq:deltaVapprox}
\Dv \equiv \dfrac{k v}{a H}\,,
\end{equation}
where $v$ is the Fourier transform of the radial baryon
velocity. Assuming that at EoR we can neglect the time evolution of
the gravitational potentials~\cite{Bharadwaj:2004nr}, one can further
approximate $\Dv \simeq -\Db$.

\subsubsection{Three-dimensional power spectrum}
\label{sec:three-dimens-power}

The three-dimensional power spectrum of the brightness temperature
fluctuations, $\PTb$, is defined via
\begin{equation}
\label{eq:Pkdef}
  \mean{\delta \Tb (\bk) \delta \Tb (\bk')} = (2 \pi)^3 \delta^3 (\bk
  - \bk') \PTb(\bk)\,,
\end{equation} 
where the brackets denote the ensemble average. One can also define
\begin{equation}
\label{eq:calPkdef}
\calPk(\bk) \equiv \dfrac{k^3}{2\pi^2} \Pk(\bk)\,,
\end{equation}
such that the isotropic real space variance simplifies to
\begin{equation}
\label{eq:varTb}
\mean{\delta \Tb^2(x)} = \int{\ud \ln k \, \calPk(k)}.
\end{equation}
Using Eq.~(\ref{eq:deltaTbhot}), the power spectrum reduces
to~\cite{Mao:2008ug}
\begin{equation}
\label{eq:PTb}
\begin{aligned}
    \PTb(\bk) & = \ue^{-2 \tauC} \Tbtilde^2 \left\{\xH^2 \Pb(k) +
    \xI^2 \Pii(k) - 2\xH \xI \Pib(k) \right. \\
    & +  \left. 2 \mu^2 \left[\xH \xI \Piv(k) -
      \xH^2 \Pbv(k) \right] + \mu^4 \xH^2 \Pv(k) \right\},
\end{aligned}
\end{equation}
where the (cross) power spectra are defined as in Eq.~(\ref{eq:Pkdef})
between the perturbation variables $\Db$, $\Dv$ and $\DxI$. The key
point of Eq.~(\ref{eq:PTb}) is that the $\mu^4$ component depends on
the baryonic matter power spectrum only since $\Pv \simeq \Pb$, which
is completely fixed by cosmology. On the contrary, the power spectra
$\Pii(k)$, $\Piv(k)$ and $\Pib(k)$ are reionization dependent and
considered as nuisances. As seen in Eq.~(\ref{eq:PTb}), they appear
only with an angular dependence in $\mu^0$ or $\mu^2$, but not in
$\mu^4$. This property is expected to play a crucial role in the
separation of the cosmological signal from those astrophysical
contaminants~\cite{Barkana:2004zy, Barkana:2004vb}. Following these
references, we can therefore define
\begin{equation}
\label{eq:P024}
\begin{aligned}
\Pzero(k) &\equiv \xH^2 \Pb(k) + \xI^2 \Pii - 2\xH \xI \Pib(k),\\
\Ptwo(k) & \equiv 2 \left[\xH\xI \Piv(k) - \xH^2 \Pvb(k) \right], \\
\Pfour(k) & \equiv \xH^2 \Pv(k),
\end{aligned}
\end{equation}
such that Eq.~(\ref{eq:PTb}) simplifies
\begin{equation}
\label{eq:PTbdecomp}
\PTb(\bk) = \ue^{-2\tauC} \Tbtilde^2 \left[\Pzero(k) + \mu^2 \Ptwo(k) + \mu^4
\Pfour(k) \right].
\end{equation}

Let us emphasize that the power spectrum $\PTb(\bk)$ is not directly
accessible to 21-cm experiments, i.e. one cannot directly measure the
comoving position $\br$ of the signal, Fourier dual of the wave vector
$\bk$. Any experiment can however determine angular separations
$\bskyang $ in the sky, and signal frequency differences $\Freq$
around any 21-cm line of redshift $z$. In this paper, we consider a
FFTT which is designed to map a hemisphere. Following
Ref.~\cite{Mao:2008ug}, one can divide the sky into small patches for
which the flat sky approximation is valid. In this limit, $\bskyang$
and $\Freq$ in each patch are directly proportional to the variations
of the comoving distance's components perpendicular $\brbot$ and
parallel $\rpara$ to the line-of-sight with respect to the central
redshift $z$,
\begin{equation}
\label{eq:thetaf}
 \bskyang = \dfrac{\brbot}{\Dang(z)}\,, \qquad \Freq =
   \dfrac{\rpara}{y(z )}\,.
\end{equation}
Here $\Dang$ stands for the angular comoving distance, given in a flat
Universe by
\begin{equation}
  \Dang(z) = \int_0 ^z \frac{1}{H(z')} \dd z'\,,
\end{equation}
and $y(z)$ is the conversion factor between comoving distances and
frequency intervals,
\begin{equation}
  y(z) = \frac{\lambdasf (1+z)^2 }{H(z)}\,.
\end{equation}
In Eq.~(\ref{eq:deltaTb}), the 21-cm brightness temperature
fluctuations have been obtained in terms of $\bk$ while it would be
more convenient to describe it in terms of Fourier duals of the
frequency differences and of the angular separations denoted as
$\upara$ and $\bubot$, respectively.  Their relation to the comoving
wave vector $\bk$ reads
\begin{equation}
  \bubot = \Dang(z) \bkbot, \qquad \upara = y(z) \kpara.
\end{equation}
As a consequence, the observable three-dimensional power spectrum of
the 21-cm brightness temperature is measured in $\bu$ space and reads
\begin{equation}
\label{eq:PktoPu}
  \PTb(\bu) = \dfrac{\PTb[\bk(\bu)]}{\Dang^2(z) y(z)}\,.
\end{equation}

\subsubsection{Angular power spectrum}
\label{sec:angpower}

Out of the flat sky approximation, one has to derive the angular power
spectrum of the 21-cm brightness temperature for each observed
redshift $z$. From an infinitely thin shell at redshift $z$,
corresponding to the observed 21-cm energy $\epsilon(z)$, the angular
power spectrum is given by~\cite{Lewis:2007kz}
\begin{equation}
  C_\ell(z) =  \dfrac{2}{\pi}  \int F_\ell(\epsilon,k) F_\ell(\epsilon,k)
  \,k^2 \ud k ,
  \label{eq:Cl21cm}
\end{equation}
where $F_\ell(\epsilon,k)$ are the multipole components today of the
21-cm photon distribution function. Their expressions have been derived
in Ref.~\cite{Lewis:2007kz} by solving the Boltzmann equation during
the dark ages for the perturbed photon distribution function (see
Appendix~\ref{sec:boltz}). At EoR, these equations remain unchanged
but the fractional fluctuations $\DHI$ appearing in
Eq.~(\ref{eq:fl21cm}) are now given by Eq.~(\ref{eq:DHI}). The
monopole source is still defined as in Eq.~(\ref{eq:Dmono}) with a
background spin temperature that can, however, be driven by the new
ionizing sources.

One can recover the usual approximated results by keeping only the
monopole and redshift distortion terms. In that case, the angular
power spectrum simplifies to
\begin{equation}
  \label{eq:Clapprox}
\begin{aligned}
  C_\ell(z) & \simeq 4\pi e^{-2\tauC(z)} \Tbtilde^2 \int \dd \ln k \bigg \{
  \calPzero(k) [j_l(\Delta\etaE)]^2 \\
  &  - \calPtwo(k) j_l(k\Delta\etaE) j_l''(k
  \Delta\etaE) + \calPfour(k) [j_l''(k\Delta\etaE]^2 \bigg \},
\end{aligned}
\end{equation}
which is the equivalent of Eq.~(\ref{eq:PTbdecomp}) on the full sky.
In the following, the full sky results are derived by using a modified
version of the $\CAMB$ code~\cite{Lewis:1999bs, Lewis:2002ah,
  Lewis:2007kz} in which we have implemented the above-mentioned
extra sources associated with the EoR. This allows to integrate the
complete equations of motion (see Appendix~\ref{sec:boltz}) under a
given model of reionization which is discussed in the next section. In
addition, we consider the same finite width Gaussian window functions
in frequency as in Ref.~\cite{Lewis:2007kz}, such that the observed
angular power spectrum is actually given by their convolution with
Eq.~(\ref{eq:Cl21cm}). Finally, we have cross-checked our computations
by comparing the modified $\CAMB$ results with an independent code
directly integrating Eq.~(\ref{eq:Clapprox}). As expected, both
spectra match at small angular scales and for narrow window
function (see Fig.~\ref{fig:thpower}).

\subsection{Reionization model}
\label{sec:eormodel}

\subsubsection{Background}

In order to model the EoR, we consider a toy model similar to the one
used for CMB data analysis~\cite{Komatsu:2010fb}, namely a smooth
transition from the dark ages ionized fraction to complete
reionization. A similar model has also been considered for 21-cm
analysis of the global reionization signal in
Ref.~\cite{Pritchard:2010pa, Morandi:2011hn}. CMB polarization
measurements allow to infer the value of $\tau$, the Compton optical
depth to reionization defined as
\begin{equation}
\label{eq:taureion}
\tau = \int_0^{\eta_0} a(\eta) \nelec(\eta) \sigmaT \ud \eta\,.
\end{equation}
In this expression, $\sigmaT$ is the Thomson scattering cross section
while $\nelec$ stands for the number density of free electrons. As
explicit in Eqs.~(\ref{eq:PTbdecomp}) and~(\ref{eq:Clapprox}), Thomson
scattering at EoR affects the overall amplitude of the 21-cm power
spectra through $\tauC(z)$, the optical depth to the redshift of
observation. Let us stress that the latter is defined as in
Eq.~(\ref{eq:taureion}) but starting at the conformal time
$\etaE$. Equations~(\ref{eq:PTbdecomp}) and~(\ref{eq:Clapprox}) make
clear that, already at the background level, the 21-cm signal at EoR
is strongly correlated with the total optical depth $\tau$ that can be
measured in CMB experiments. In order to make the dependency explicit,
we have used and extended the reionization model used in
$\CAMB$~\cite{Lewis:1999bs}. In this model, and denoting by $\xe
\equiv \nelec/(\nH+\nI)$ the total ionized fraction\footnote{It can
  therefore be greater than one when helium becomes ionized.},
reionization occurs according to
\begin{equation}
\label{eq:xereio}
\begin{aligned}
\xe & = \dfrac{1+\fnHe}{2} \left\{1 + \tanh\left[\dfrac{(1+\zreio)^{3/2} -
    (1+z)^{3/2}}{\DeltaEoR} \right] \right\} \\
& + \dfrac{\fnHe}{2} \left[ 1+ \tanh\left( \dfrac{\zHe-z}{\DeltaHe}
  \right) \right] +\left.\xe\right|_{\urec},
\end{aligned}
\end{equation}
where $\fnHe\simeq 0.08$ is the helium density fraction, assumed to
become firstly ionized at the same time as
hydrogen\footnote{$\fnHe=\dfrac{\mH}{\mHe} \dfrac{Y}{1-Y}$.} and
doubly ionized at $z=\zHe$. The parameter $\DeltaEoR$ fixes the
duration of reionization while $\zreio$ refers to its central
redshift. The second term encodes the complete helium ionization and
remains always very small. It has been kept only for consistency with
previous CMB analysis with the same fixed values of $\zHe=3.5$ and
$\DeltaHe=0.5$. The last term is a constant encoding the residual
ionized fraction from recombination which does not exceed a few
thousandths. From Eq.~(\ref{eq:xereio}), one immediately gets the
associated optical depth by using Eq.~(\ref{eq:taureion}) and
conversely. The choice of the power $3/2$ in Eq.~(\ref{eq:xereio}) is
made such that for all $\DeltaEoR$, the corresponding value of $\tau$
remains equal to the one that would be associated with an
instantaneous reionization at $z=\zreio$ (see
Ref.~\cite{Lewis:1999bs}). It can be convenient to express $\DeltaEoR$
in terms of redshift duration and we will use in the following the
parameter $\Deltaz$ defined by
\begin{equation}
\Deltaz \equiv \dfrac{2}{3} \dfrac{\DeltaEoR}{\sqrt{1+ \zreio}}\,.
\end{equation}

For our purpose, Eq.~(\ref{eq:xereio}) already fixes the evolution of
the hydrogen neutral and ionized fraction, respectively, as
\begin{equation}
\label{eq:xHreio}
\begin{aligned}
  \xH &= 1 - \xI,\\
  \xI & = \left. \xI \right|_{\urec} \\ + &
  \dfrac{1-\left. \xI\right|_{\urec} }{2} \left\{1 + \tanh \left[
      \dfrac{(1+\zreio)^{3/2} - (1+z)^{3/2}}{\DeltaEoR} \right]
  \right\}.
\end{aligned}
\end{equation}
Concerning the spin temperature, since it is expected to be physically
driven by ionizing sources, we have chosen for simplicity a similar
evolution:
\begin{equation}
\label{eq:Tspinreio}
\begin{aligned}
\Tspin(z) & = \TspinRec + \dfrac{\TspinMax-\TspinRec}{2} \\
& \times \left\{1 + \tanh \left[ \dfrac{(1+\zreio)^{3/2} -
    (1+z)^{3/2}}{\DeltaEoR} \right] \right\}.
\end{aligned}
\end{equation}
The quantity $\TspinRec$ is the spin temperature value during the dark
ages, just before reionization starts, and it is completely fixed by
the cosmological model. Let us emphasize that in a more realistic
situation the redshift functional dependence of $\Tspin$ is not the
same as the ionized fraction, especially during the first stages of
the reionization processes~\cite{Furlanetto:2006jb,
  Pritchard:2011xb}. For our purpose, Eq.~(\ref{eq:Tspinreio})
introduces only one extra parameter, $\TspinMax$, which gives the
maximum spin temperature value once the Universe is completely
reionized. Unless specified, we assume in the following the fiducial
$\TspinMax = 10000\,\K$.

\begin{figure}
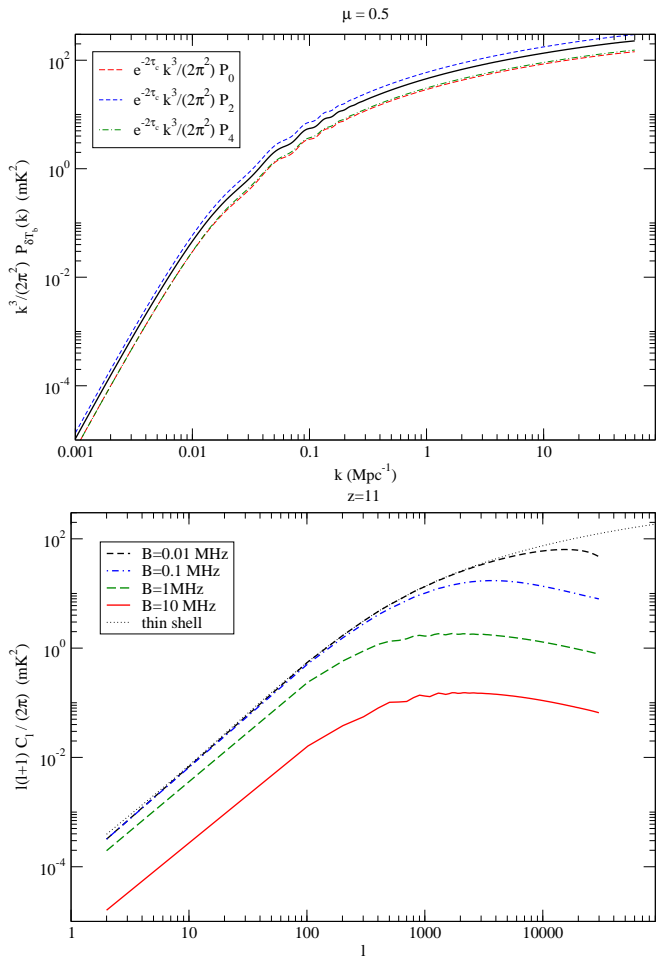

\begin{center}
 \includegraphics[width=\columnwidth]{pkthspectra}
 \includegraphics[width=\columnwidth]{clsthspectra}
 \caption{Three-dimensional power spectra $\calPTb(k)$ (top) and
   angular power spectrum $C_\ell$ (bottom) for an observation
   redshift $z=11$ and assuming no foregrounds. The cosmological
   parameters are the ones of Sec.~\ref{sec:eormodel} ($\tau=0.088$,
   $\zreio \simeq 10.54$, $\xH\simeq 0.87$, $\Tspin \simeq 1360\,\K$). The
   $C_\ell$ are convolved with Gaussian window functions whose
   standard deviation is given by the bandwidth $\bwidth$. Large
   bandwidths damp the overall power but render the acoustic
   oscillations visible~\cite{Lewis:2007kz}. The thin shell limit has
   been obtained by integrating Eq.~(\ref{eq:Clapprox}).}
    \label{fig:thpower}
\end{center}
\end{figure}

Figure~\ref{fig:thpower} shows the theoretical three-dimensional and
angular power spectra computed at $z=11$ for the background
cosmological and reionization models just described. The top curve in
the bottom panel has been obtained by a direct integration of
Eq.~(\ref{eq:Clapprox}), i.e. including only monopole and redshift
distortions. As expected, it matches in shape and amplitude with
three-dimensional power spectrum (top panel). The other angular
spectra have been computed with our modified version of the $\CAMB$
code and are convolved in frequency with Gaussian window function of
bandwidth $\bwidth$. As discussed in Ref.~\cite{Lewis:2007kz}, larger
values of the bandwidth damps the overall power but allows the
sampling of acoustic oscillations. In the limit $\bwidth \rightarrow
0$, one recovers Eq.~(\ref{eq:Clapprox}).

\subsubsection{Ionizing sources}
\label{sec:fore}
At the perturbative level, the shape and time evolution of $\Pib$,
$\Pvb$ and $\Pii$ are expected to depend on the reionization
details. Following Ref.~\cite{Mao:2008ug}, we consider here only the
spectra associated with the fluctuations in the ionized fraction and
adopt a phenomenological parametrization of these power spectra as
\begin{equation}
\label{eq:reionizedpower}
\begin{aligned}
\Pii(k) & = \Nii \left[1 + \alphaii\, k \Rii + (k \Rii)^2
\right]^{-\gammaii/2} \Pb(k),\\
\Pib(k) & = \Nib \exp\left[-\alphaib \, k \Rib - (k \Rib)^2 \right]
\Pb(k)\,
\end{aligned}
\end{equation}
where $\Nii$, $\Nib$ are normalization constants and $\alphaii$,
$\alphaib$, $\gammaii$, $\Rii$ and $\Rib$ are model parameters to be
adjusted to match numerical simulations~\cite{McQuinn:2007dy,
  Zahn:2010yw}. Since these spectra are nuisance, we moreover
approximate $\Piv(k) \simeq -\Pib(k)$ from
Eq.~(\ref{eq:deltaVapprox}).  Taking the fiducial cosmological
parameter values $\OmegaB h^2= 0.023$, $\OmegaDM h^2= 0.115$, $h=
0.71$ and $\tau=0.088$, the reionization model of the previous section
gives $\zreio =10.536$ at which the hydrogen neutral fraction
$\xH(\zreio)=0.5$.

\begin{table}
\begin{tabular}{|c|c|c|c|c|c|c|c|c|c|}
\hline
$z$ & $\xH$ & $\Nii$ & $\Rii\, (\Mpc)$ & $\alphaii$ & $\gammaii$ & $\Nib$ &
$\Rib\, (\Mpc)$ & $\alphaib$ \\
\hline
11  &  0.87  &   16.0   &   1.3  &  -1.4  &   0.5  & 3.96  & 0.57  & -0.27\\
\hline     
\end{tabular}
\caption{Parameter values for the nuisance ionization power spectra
  at $z=11$, linearly interpolated from Ref.~\cite{Mao:2008ug}.}
\label{tab:ioparams}
\end{table}

In Table~\ref{tab:ioparams}, we give the parameter values associated
with the reionization power spectra of
Refs.~\cite{McQuinn:2007dy,Mao:2008ug} at $z=11$. These values will be
used in the following to test the robustness of the parameter
estimation with respect to the presence of extra ionizing
sources. Although not rigorously correct, we assume that the redshift
dependence of the ionizing power spectra is completely given by the
ionized fractions $\xH(z)$ and $\xI(z)$, i.e. that the parameters of
Table~\ref{tab:ioparams} remain constant during the beginning of the
EoR. Notice that the ionized fraction dependency is not included in
$\Nii$ and $\Nib$ such that the time evolution of the ionizing power
spectra is well behaved, i.e. vanishes as soon as $\xI \rightarrow
0$.

\begin{figure}
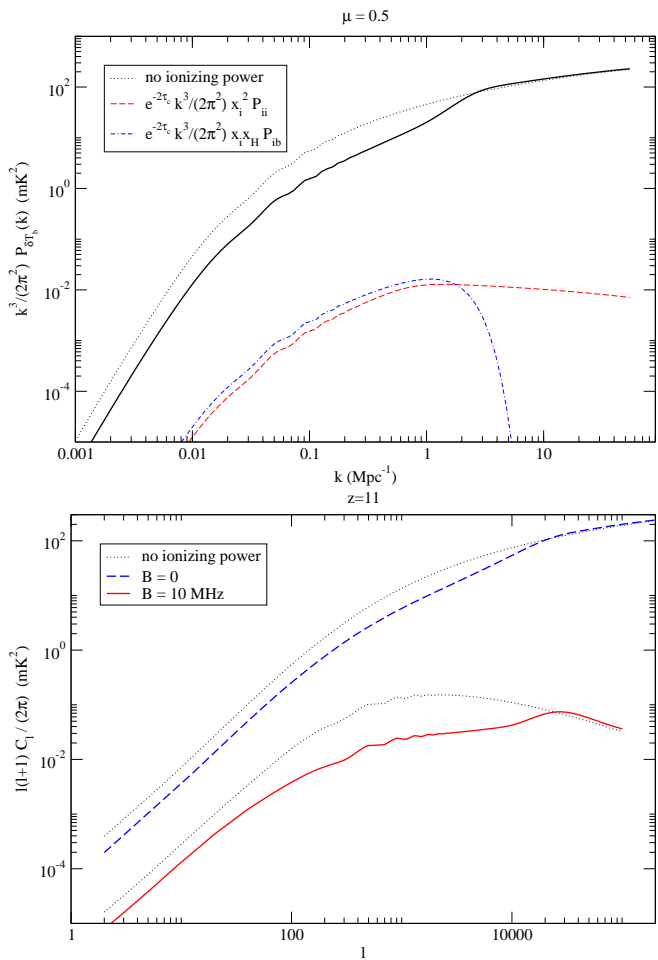

\begin{center}
 \includegraphics[width=\columnwidth]{pkiospectra}
 \includegraphics[width=\columnwidth]{clsiospectra}
 \caption{Effects of the reionization power spectra at $z=11$ (see
   Table~\ref{tab:ioparams}) on the three-dimensional power spectrum
   $\calPTb$ (top). The bottom panel shows the reionization induced
   deviations on the angular power spectrum $C_\ell$, both for the
   thin shell approximation and a window function of bandwidth
   $\bwidth=10\,\MHz$.}
    \label{fig:iopower}
\end{center}
\end{figure}

In Fig.~\ref{fig:iopower}, we have plotted these spectra for $z=11$,
as well as their effects on the expected three-dimensional and angular
power spectrum. For this redshift, $\xH \xI \Pib$ dominates and
reduces the overall power whereas at low redshift we would find extra
power coming for the domination of $\xI^2 \Pii$. Finally, let us
emphasize that we have not included, for simplicity, the contributions
coming from the spin and gas temperature fluctuations that would be
associated with the ionizing sources. However, such an improvement
could be made in a similar way.

\subsection{Fast Fourier Transform Telescope}
\label{sec:fftt}

In this section we give the ideal characteristics chosen for the
omniscope used in the following forecasts. The FFTT we are
considering is in between the original design introduced in
Ref.~\cite{Tegmark:2008au} and a very futuristic concept used in
Ref.~\cite{Gordon:2009wx} for probing the dark ages. Detecting signals
from the EoR does not require gigantic array as the redshifted 21-cm
wavelength from $\zreio$ is about two meters. However, pushing up
the scale of the original FFTT concept from $1\,\km$ to $10\,\km$
would allow to probe the end of the dark ages, i.e. up to redshift
$20$--$30$. In such a design, the sensitivity to dark age signals is
quite low and  forecasts suggest that this would not be competitive
with CMB (see Sec.~\ref{sec:dagesfc}). However, including those
redshifts into a multiredshift analysis allows to slice in details
the transition to the EoR, and as discussed in Sec.~\ref{sec:eorfc},
allows a good inference of the background reionization properties such
as duration or spin temperature evolution.

\subsubsection{Beam, noise and ionosphere}

We consider a squared FFTT design of $D=10\,\km$ sides and constituted
of ten million antennas, i.e., one every
$3\,\meters$~\cite{Tegmark:2008au}. For simplicity, the beam is
approximated by a Gaussian whose full width at half maximum is fixed
to $\fwhm= 0.89 \lambda/D$ to match the actual square aperture. In
terms of multipole moments, defining
\begin{equation}
\lbeam \equiv \dfrac{4 \sqrt{\ln 2} }{\fwhm}\,
\end{equation}
the beam function (normalized to unity for $\ell=0$) reads
\begin{equation}
\label{eq:gbeam}
B_\ell = \ue^{-\ell (\ell+1)/\lbeam^2}\,.
\end{equation}
For a system antenna temperature given by $\Tsys$ the induced flux
density fluctuations of one receptor of collecting area $\Aone$ have a
root mean square given by~\cite{2001:Thompson, 2009:Wilson}
\begin{equation}
\delta \Snu = \dfrac{2 \Tsys}{\Aone \sqrt{\texp \bwidth}}\,.
\end{equation}
Here $\texp$ stands for the exposure time, $\bwidth$ is the frequency
bandwidth and the flux density is related to the brightness
temperature by $\Snu = 2 \Tb/\Aone$. For aperture synthesis against
$\Na$ antennas, these fluctuations are reduced by a factor
$\sqrt{\Na(\Na-1)/2} \simeq \Na/\sqrt{2}$ corresponding to the square
root of the number of simultaneous pairwise correlations. As a
result, the noise fluctuations in the brightness temperature for an
array of area $D^2$ read
\begin{equation}
  \dTn = \dfrac{1}{\sqrt{2}} D^2 \dfrac{\delta \Snu}{\Na} \simeq
  \dfrac{\Tsys}{\fcover \sqrt{\texp \bwidth}}\,,
\end{equation}
up to an order unity factor. The covering factor $\fcover \equiv \Na
\Aone /D^2 = \Aeff/D^2$. For a sky map, $\dTn^2$ gives the expected
noise variance in each pixel subtending a solid angle $\dOmega \simeq
(\lambda/D)^2$. The angular noise power is thus given
by~\cite{Zaldarriaga:2003du}
\begin{equation}
\label{eq:varTn}
\dTn^2 \dOmega \simeq \dfrac{\lambda^2}{D^2
  \fcover^2} \dfrac{\Tsys^2}{\bwidth \texp} = \dfrac{\lambda^2}{\Aeff
  \fcover} \dfrac{\Tsys^2}{\bwidth \texp}\,.
\end{equation}

Depending on the scanning strategy, a telescope may eventually map a
solid angle $\OmegaSky=4 \pi \fsky$ larger than its instantaneous
field of view $\OmegaFov$. This may be taken into account by
renormalizing the exposure time by a factor
$\OmegaFov/\OmegaSky$~\cite{Tegmark:2008au}. The angular noise power
finally reads
\begin{equation}
\label{eq:Csys}
\Csys \simeq \dfrac{\lambda^2}{D^2 \fcover^2}
\dfrac{\Tsys^2}{\bwidth \texp} \dfrac{4 \pi \fsky}{\OmegaFov}\,.
\end{equation}
It is well known from CMB data analysis that, for uncorrelated signal
and noise, the root mean square fluctuations for the angular power
spectrum are given by~\cite{Knox:1995dq, 2008:durrer}
\begin{equation}
  \delta C_\ell = \sqrt{\dfrac{2}{2 \ell+1}} \left(C_\ell +
    \dfrac{\Csys}{B_\ell^2} \right).
\end{equation}
The beam effects can thus be included by defining the noise angular
power spectrum to be
\begin{equation}
\label{eq:Cnoise}
\Cnoise_\ell \equiv \dfrac{\Csys}{B_\ell^2}\,.
\end{equation}

These results also apply to the three-dimensional power
spectrum. Starting from Eq.~(\ref{eq:varTn}), the corresponding noise
spectrum $\Pnoise$ reads
\begin{equation}
\label{eq:Pnoisek}
  \Pnoise(\bk) \simeq \dTn^2 \dV = \Csys \Dang^2 y \bwidth,
\end{equation}
where $\dV = \Dang^2 y \dOmega \bwidth$ is the comoving volume of the
three-dimensional pixel intercepting the solid angle $\dOmega$ and for
a redshift resolution associated with the bandwidth $\bwidth$ [see
Eq.~(\ref{eq:thetaf})]. From Eq.~(\ref{eq:PktoPu}), one gets the noise
power spectrum in the $\bu$ variable
\begin{equation}
\label{eq:Pnoiseu}
\Pnoise(\bu) \simeq \dfrac{\lambda^2}{D^2 \fcover^2}
  \dfrac{\Tsys^2}{\texp} \dfrac{4 \pi \fsky}{\OmegaFov}\,.
\end{equation}
Beam effects at small scales can be incorporated by multiplying this
expression by $\exp{( \bubot^2/\lbeam^2)}$.

Finally, ionospheric absorption becomes important for all frequencies
approaching the plasma frequency $\nup \simeq 12\,\MHz$ such that
atmosphere is opaque to 21-cm signal coming from $z \gtrsim 100$. For
frequencies $\nu \gg \nup$ ($z \ll 100$), the ionospheric absorption
coefficient $\Aiono$ on the flux density can be approximated
by~\cite{2001:Thompson},
\begin{equation}
\label{eq:ionoabs}
10 \log(\Aiono) \lesssim 0.5\,\left( \dfrac{100\,\MHz}{\nu} \right)^2.
\end{equation}
For forecasting, instead of applying this damping directly to the
expected signal, we prefer to rescale the noise power by multiplying
Eqs.~(\ref{eq:Cnoise}) and (\ref{eq:Pnoiseu}) by $\Aiono^2$.

\subsubsection{Mock power spectra}

\begin{figure}
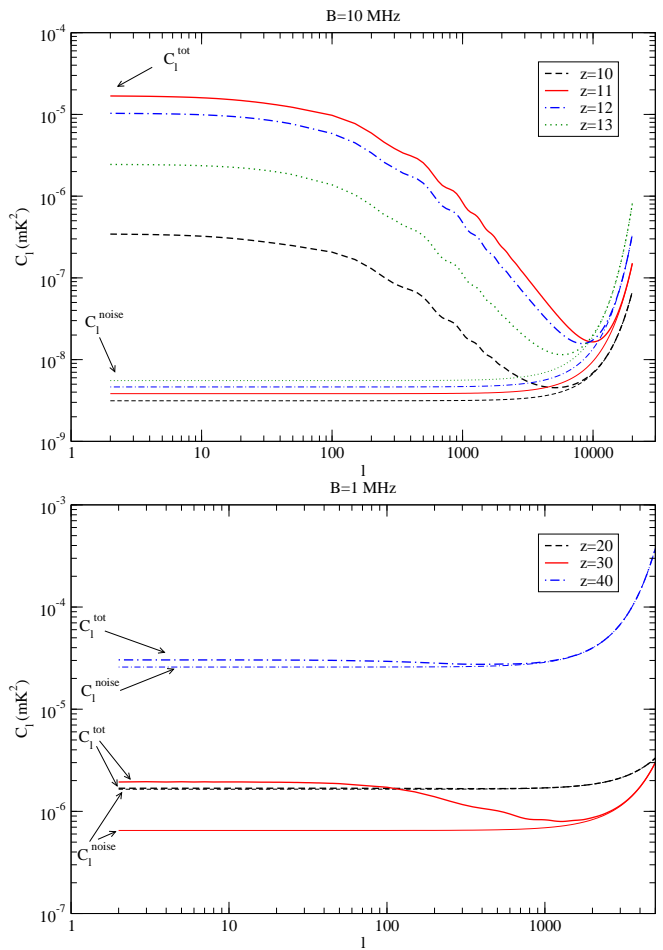

\begin{center}
 \includegraphics[width=\columnwidth]{fftt_10MHz_10k}
 \includegraphics[width=\columnwidth]{fftt_1MHz_10k}
 \caption{Mock angular total power spectra $C_\ell+\Cnoise_\ell$ and
   $\Cnoise_\ell$ in $\mK^2$ for various observation redshifts around
   reionization (top) and at the end of the dark ages (bottom). The
   cosmological parameters are the ones of Sec.~\ref{sec:eormodel}
   ($\tau=0.088$, $\zreio \simeq 10.5$). The window functions are
   Gaussian in frequency space with a respective bandwidth of
   $\bwidth=10\,\MHz$ (top) and $\bwidth=1\,\MHz$ (bottom). The FFTT
   design is for $D=10\,\km$ and allows probing the dark ages for $z
   \lesssim 30$, above which ionospheric absorption damps the signal
   (see $z=40$).}
    \label{fig:sncls}
\end{center}
\end{figure}

The fiducial design considered in the following has a size of
$D=10\,\km$, with a covering factor of $\fcover=0.1$ and an exposure
time of one year, $\texp \simeq 3.1 \times 10^7\,\sec$. The system
temperature is chosen larger than its typical value, $\Tsys =
400\,\K$, to account for some extra noise. Notice that foregrounds
removal is generically scale dependent and cannot be simply accounted
by a rescaling of $\Tsys$~\cite{Mitchell:2010cc, Datta:2010pk,
  Morales:2012kf}. However, as already mentioned, we are here
considering an ideal situation and we do not intend to model effects
from foregrounds' residuals. For omniscope, the instantaneous field of
view and sky coverage have been chosen equal to the visible sky
$\OmegaSky=\OmegaFov = 2\pi$, i.e. $\fsky=0.5$. In
Fig.~\ref{fig:sncls}, we have plotted the angular power spectrum
expected for such a design at various observation redshifts $z$.

\section{Forecasts on cosmological and reionization parameters}
\label{sec:reiocast}

In order to minimize the parameter space dimensions, we consider a
simple primordial power spectrum which is a power law with
unobservable tensor modes. As shown below, even for vanishing
ionization power spectra, FFTT alone does not constrain well the
optical depth $\tau$ damaging the expected bounds on the background
cosmological parameters. As one of the unconstrained parameters is the
baryon density $\OmegaB h^2$, the situation can be greatly improved by
combining FFTT with CMB data. Taking four redshift slices for the FFTT
and Planck-like typical CMB data, we forecast the expected errors on
the cosmological parameters and show that all the parameters entering
the reionization models can be inferred. At last, the impact of the
ionizing power spectra is discussed.

\subsection{Methodology}

In the following, we use both Fisher matrix analysis for the
three-dimensional power spectra, and MCMC exploration of the parameter
space for the angular 21-cm and CMB power spectra. This allows us to
discuss how reliable are the expected constraints in presence of
degeneracies.

Unless specified, our fiducial cosmological model is the one
introduced in Sec.~\ref{sec:models} complemented with a scalar power
law primordial power spectrum $\calPs(k) = \As (k/\kstar)^{\nS-1}$,
with a pivot scale $\kstar=0.05\,\Mpc^{-1}$. The fiducial spectral
index has been fixed to $\nS=0.973$ and the amplitude to
$\ln(10^{10}\As) = 3.16$. For convenience, we have sampled over the
CMB parameter $\theta$, the angular scale of the sound horizon at last
scattering, instead of the Hubble parameter $H_0$. This facilitates
joint CMB and 21-cm analysis discussed later and we have checked that
this does not affect significantly the forecasts for 21-cm alone.

\subsubsection{Fisher matrix analysis} \label{sec:Fisher}

Following the standard approach, given a fiducial set of cosmological
and reionization parameters $\{\fidlambda_a \}$, the expected errors
on FFTT measurements of the power spectrum $\PTb(\bu)$ can be
evaluated by using the Fisher matrix formalism~\cite{Tegmark:1996bz}.
Let us assume that we can compute the likelihood ${\cal L}$, i.e. the
expected distribution of the data given a certain model. The Fisher
matrix approach assumes that the expected behaviour of the likelihood
near the maximum is Gaussian such that its curvature alone allows to
estimate errors on the parameters. The Fisher information matrix is
defined as
\begin{equation}
\bfish_{ab} =-\left\langle \frac{\partial^2\ln{\cal
    L}}{\partial \lambda_a \partial \lambda_b}\right\rangle,
\end{equation}
where the brackets denote an ensemble average. When all the parameters
are estimated simultaneously, the marginalized error on a given
parameter reads $\sigma_{\lambda_a}\geq
\sqrt{(\bfish^{-1})_{aa}}$. The Cram\'er-Rao inequality (``$\geq$'')
emphasizes the fact that the Fisher matrix approach always gives a
local optimistic estimate of the errors. It becomes an equality when
the likelihood is a Gaussian around its maximum. In the following, we
estimate the expected variance on the measured parameters $\lambda_a$
from the diagonal elements of the covariance matrix, i.e. from the
inverse Fisher matrix elements
\begin{equation}
  \mean{\delta \lambda_a^2} = \left(\bfish^{-1}\right)_{aa}.
\end{equation}

For the purpose of the Fisher matrix analysis, as already mentioned in
Sec.~\ref{sec:three-dimens-power}, one can divide the sky into small
patches for which the flat sky approximation is valid.  Given an
angular patch $\skyang$ and a frequency bin of size $\Freq$, the
probed comoving volume is
\begin{equation}
  \label{eq:V}
  V = \skyang^2 \Dang^2 \Freq y\,.
\end{equation} 
Following~\cite{Mao:2008ug}, one can consider step sizes
in the dual $\bu$ space as follows:
\begin{equation}
  \label{eq:ustep}
  \delta \ubot  =  \dfrac{2 \pi}{\skyang}\,, \quad     
    \delta \upara =   \frac{2 \pi}{\Freq}\,,
\end{equation}
where $\skyang$ is taken to be lower than typically $1\, \rad$ and
$\Freq$ is set by the frequency size of the redshift bin. Assuming
that $\PTb(\bu)$ is Gaussian distributed, we can approximate the
Fisher matrix by
\begin{equation}
  \bfish_{ab} = \dfrac{1}{2} \sum_{\upara,\ubot} \dfrac{\Nc}{\left[\PTb(\bu) +
      \Pnoise \right]^2}
  \dfrac{\partial \PTb(\bu)}{\partial\lambda_a} \dfrac{\partial
    \PTb(\bu)}{\partial \lambda_b}\,,
  \label{eq:fish}
\end{equation}
where
\begin{equation}
  \Nc = \frac{4 \pi \fsky}{\skyang^2} 2 \pi \kbot \delta \kbot \delta \kpara 
  \dfrac{V}{(2\pi)^3}\,,
\end{equation}
is the number of independent cells probed\footnote{The factor $1/2$ in
  Eq.~(\ref{eq:fish}) takes into account the fact that $\vec k$ and
  $-\vec k$ are not independent.}  for a given value of $\bk$, or
equivalently of $\bu$. Notice that given Eqs.~(\ref{eq:V}) and
(\ref{eq:ustep}), $\Nc$ reduces to
\begin{equation}
  \Nc=  \dfrac{4 \pi \fsky}{\skyang^2} \dfrac{2 \pi
    \kbot}{\delta \kbot}\,.
  \label{eq:Nc}
\end{equation}

In the following, we will be interested in comparing the results of
the Fisher analysis on the three-dimensional power spectrum
$\PTb(\bu)$ to the results of a MCMC analysis which makes use of the
angular power spectrum $C_\ell$. From our experimental design, the
measured 21-cm power spectrum is, however, convolved in frequency by a
Gaussian window function of width $B$. As the result, and unless
specified, we will also compute the Fisher matrices of
Eq.~(\ref{eq:fish}) from the three-dimensional power spectrum at
$\kpara =0$ after a convolution with the same Gaussian window
function. Compared to Ref.~\cite{Mao:2008ug}, this allows us to
consider larger redshift bins, and thus larger bandwidths, in which
the power spectrum and background quantities are allowed to vary. In
Ref.~\cite{Mao:2008ug}, the FFTT was assumed to have a small bandwidth
of $B \simeq 0.15\,\MHz$. Around each observation redshift, the signal
on a bin of size $\delta z \simeq 0.5$ is expanded in Fourier modes
$\upara$, up to $\uparamax = 2 \pi / B$.  Within our approach, this
Fourier expansion is now replaced by a convolution of the signal with
the frequency window function associated with the mean redshift of
observation. Combining multiple redshifts is still done by adding up
their respective Fisher matrices.

Let us stress that when the bandwidth is small enough, there is no
need for the convolution. Summing the Fisher on $\upara$ running on
positive and negative values with $|\uparamax|$ set by the frequency
resolution of the experiment, i.e. the bandwidth $\bwidth$, gives the
same result than subdividing the redshift bin in an identical number
of slices to the number of parallel modes.  We have cross-checked our
method by using the FFTT specifications and reionization model of
Ref.~\cite{Mao:2008ug}, for which we recover the same results, in the
small bandwidth limit.

Concerning the transverse modes, since the volume of the experiment is
finite, only a discrete number of modes is theoretically accessible to
the analysis. In Eq.~(\ref{eq:fish}), the sum over $\ubot$ runs over
positive values from $\ubotmin=2 \pi/\skyang$, to $\ubotmax \simeq 2
\pi D / \lambda$, the angular resolution being set by the longest
baseline of the experiment, $D$.

Finally, let us also mention that when constructing the Fisher matrix,
one has to pay special attention to the numerical accuracy at which
the power spectrum derivatives have been estimated. Indeed, the
evolution of the 21-cm power spectrum with the cosmological parameters
can be highly non linear and is furthermore redshift dependent. As
discussed in Ref.~\cite{Hall:2012kg}, one has to find a good
compromise between small step sizes in the numerical evaluation of the
derivatives and the appearance of numerical noise. For each parameter,
a step size corresponding roughly to $2\%$ of the fiducial values was
found to be a good compromise.  For each redshift, the power spectrum
needs to be calculated for a large number of wavelength modes
(typically $3000$), and in order to spare computing time, it has been
interpolated using $20$ $k$ values per unit logarithmic interval. The
derivatives have been evaluated by means of a four-point method for
well-behaved parameters whereas for $\nS$, $\As$, plus the nuisance
reionization parameters $\Nii, \Rii, \alphaii, \gammaii, \Nib, \Rib$
and $\alphaib$, the derivatives have been determined
analytically. Finally, to keep the numerical noise under control, the
transfer functions have been computed using high precision settings
within $\CAMB$.

\subsubsection{Markov Chain Monte Carlo}

From the mock data associated with the FFTT design of
Sec.~\ref{sec:fftt}, forecasts can be derived through MCMC exploration
of the parameter space provided one specifies the likelihood. As for
the Fisher analysis, the mock data $\{\fidC_\ell\}$ are assumed to be
associated with a set of fiducial parameters $\{\fidlambda_a\}$. For a
full sky analysis, one can show that the sampling distribution
followed by the $C_\ell$ is a gamma
distribution~\cite{Hamimeche:2008ai}. For a cut sky, with an isotropic
beam and noise, one can use the approximated
likelihood~\cite{Bond:1998qg, Percival:2006ss}
\begin{equation}
\label{eq:like}
-2 \ln \like{D_\ell}{\fidD_\ell} = \sum_\ell \fsky (2\ell+1)
\left( \dfrac{\fidD_\ell}{D_\ell} + \ln \dfrac{D_\ell}{\fidD_\ell} -1 \right).
\end{equation}
Here $D_\ell$ denotes the theoretical angular power spectrum, plus
noise, i.e.
\begin{equation}
D_\ell = C_\ell + \dfrac{\Cnoise_\ell}{B_\ell^2}\,,
\end{equation}
associated with any tested set of parameters $\{\lambda_a\}$. The same
equation is used to define the $\fidD_\ell$ from the fiducial
$\fidC_\ell$. From our modified version of the $\CAMB$ code, and FFTT
specifications, the $D_\ell$ can be computed for any input value of
the cosmological and reionization parameters $\{\lambda_a\}$. Using
MCMC sampling with the likelihood of Eq.~(\ref{eq:like}) allows to
extract their posterior probability distribution for a given fiducial
model $\{\fidlambda_a\}$. In the following, we have used a modified
version of the public code $\COSMOMC$ coupled with our modified
$\CAMB$ code~\cite{Lewis:2002ah}.

\subsection{Single redshift from dark ages}
\label{sec:dagesfc}

\begin{figure}
\begin{center}
  \includegraphics[width=\columnwidth]{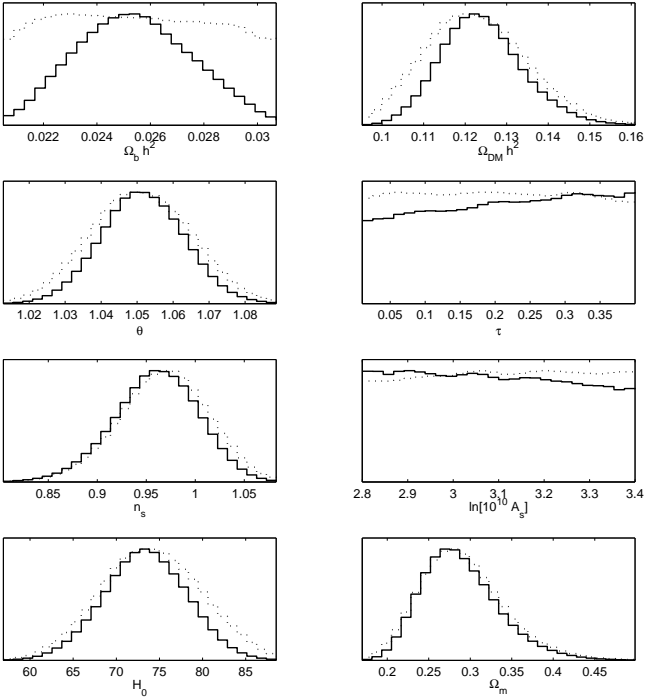}
  \caption{Expected marginalized posterior probability distributions
    (solid) and mean likelihood(dotted) for $10\,\km$ FFTT data alone,
    at $z=30$ (dark ages). The total optical depth $\tau$, primordial
    amplitude $\As$, baryon density $\OmegaB h^2$ are completely
    degenerated and not constrained (see text and Fig.~\ref{fig:z30_2D}).}
  \label{fig:z30_1D}
\end{center}
\end{figure}

In this section, the above-described MCMC analysis is applied to one
dark ages redshift $z=30$. As discussed in Sec.~\ref{sec:fftt}, the 21
cm physics during dark ages is free of reionization uncertainties, up
to the optical depth. Within our design, it has a low signal-to-noise
ratio whose only interest is to exacerbate the degeneracies induced by
the optical depth into the other cosmological parameters. In
Fig.~\ref{fig:z30_1D}, we have plotted the marginalized probability
distribution on the cosmological parameters for the same reionization
model used for CMB data analysis, namely only $\tau$ (or $\zreio$) is
varying.

\begin{figure}
\begin{center}
  \includegraphics[width=\columnwidth]{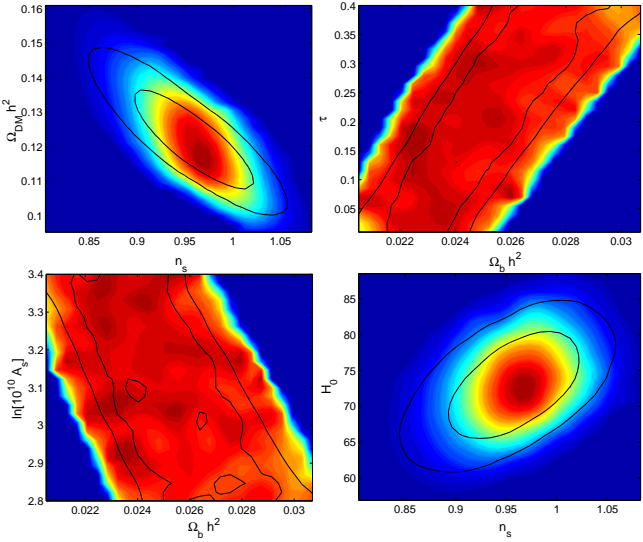}
  \caption{One- and two-sigma confidence interval of the
    two-dimensional marginalized posterior probability distributions
    (solid lines) for a $10\,\km$ FFTT design at $z=30$. The shading
    traces the two-dimensional mean likelihood.}
  \label{fig:z30_2D}
\end{center}
\end{figure}

Both the primordial power spectrum amplitude $\As$ and $\tau$ enter
into the amplitude of the 21-cm signal and one expects them to be
degenerated. As Figs.~\ref{fig:z30_1D} and \ref{fig:z30_2D} show,
these parameters are in fact mostly degenerated with $\OmegaB
h^2$. This is expected because, as can be seen in
Eq.~(\ref{eq:Tbtilde}), the baryon density also enters into the
overall amplitude of the signal. Moreover, even for a fixed $\tau$,
and fixed observation redshift $z$, $\tauC$ still inherits extra
dependencies on the other cosmological parameters from their influence
on the background evolution. As a result, all parameters end up being
more or less contaminated by the reionization parameter $\tau$, which
seriously damages the expected constraints compared to a situation in
which $\tau$ would be perfectly known. The notable exception concerns
the spectral index $\nS$, simply because the number of modes
accessible to the FFTT is large enough to detect its effect over
background degeneracies, these ones affecting all modes in a smoother
way. Nevertheless, the forecasts on $\nS$ do not compete with CMB
bounds. Let us stress that we have used only one redshift slice to
emphasize the above-mentioned degeneracies and one could add other
redshifts to improve the constraints. However, for the dark ages, the
signal dominates over the noise only around $z\simeq 30$ (see
Fig.~\ref{fig:sncls}) and the above degeneracies still remain. The
situation is better at EoR.

\subsection{Epoch of Reionization}
\label{sec:eorfc}

The fiducial model is unchanged but the observation redshifts now
probe the beginning of the EoR.

\subsubsection{Optical depth induced degeneracies for one redshift}

\label{sec:z11}

As for the dark ages, we have first performed our analysis for a
single redshift slice $z=11$, close to the maximum signal-to-noise
ratio of the $10\,\km$ FFTT and sampled at $\bwidth=10\,\MHz$. In an
optimal situation, the ionizing power spectrum can be switched off and
the background reionization model has only $\tau$ varying.

\begin{figure}
\begin{center}
  \includegraphics[width=\columnwidth]{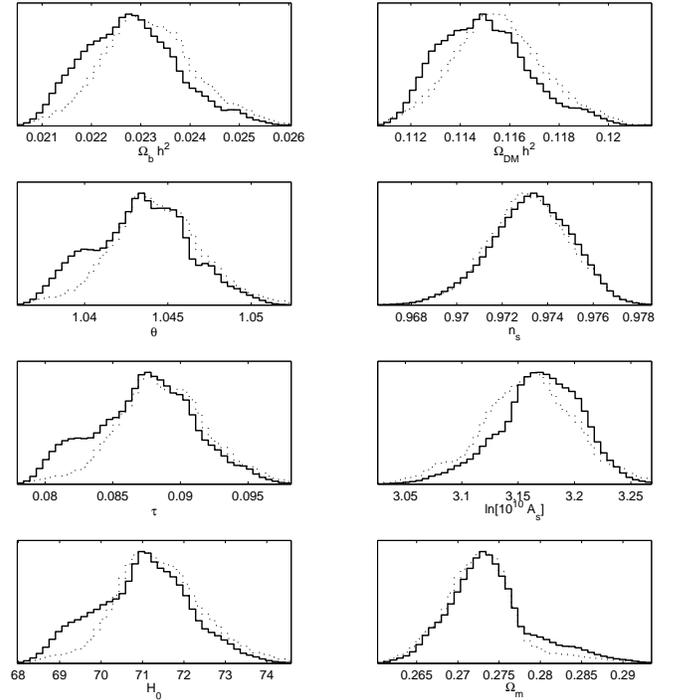}
  \caption{Expected marginalized posterior probability distributions
    (solid) and mean likelihood(dotted) from a single redshift
    observation at $z=11$ (reionization). The non-Gaussian posteriors
    and ``shoulders'' apparent for some parameters comes from the
    large correlations induced by $\tau$ (see
    Fig.~\ref{fig:z11_2D}).}
  \label{fig:z11_1D}
\end{center}
\end{figure}

\begin{figure}
\begin{center}
  \includegraphics[width=\columnwidth]{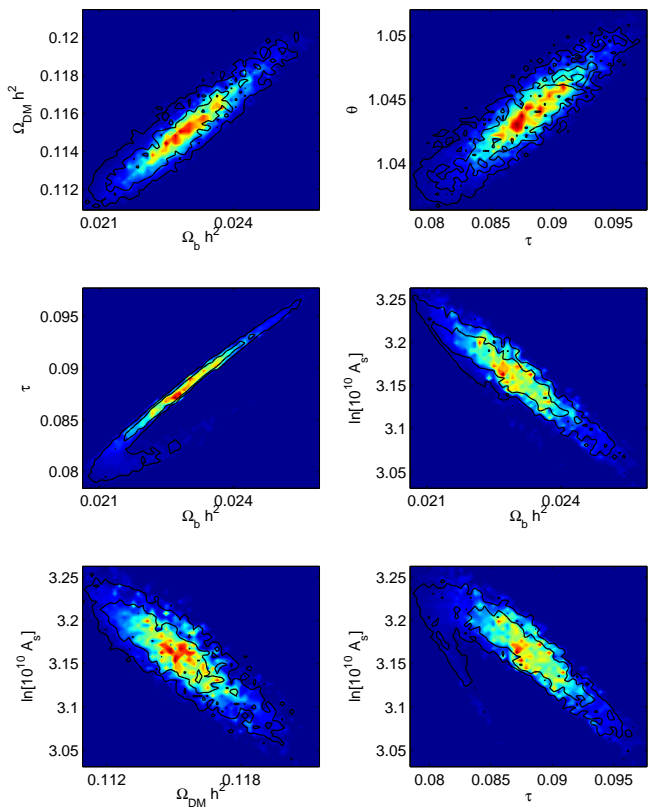}
  \caption{Two-dimensional posteriors for a single observation
    redshift at $z=11$. The solid lines represent the one and
    two-sigma confidence interval while the shading traces the
    two-dimensional mean likelihood. Notice the strong degeneracies
    induced by $\tau$ which boost the expected variance on $\OmegaB
    h^2$ for instance.}
  \label{fig:z11_2D}
\end{center}
\end{figure}

The MCMC forecasts are represented in Figs.~\ref{fig:z11_1D} and
\ref{fig:z11_2D} as one- and two-dimensional posterior distributions.

As for the dark ages, the degeneracies induced by $\tau$ are still
present and propagate to the other parameters as well. However,
$\tau$, $\As$ and $\OmegaB h^2$ are now constrained. This comes from
the signal to noise ratio at $z=11$ which is much larger than for the
dark ages, but also from the stronger effects induced by $\tau$ which
makes the degeneracies nonlinear. As seen in Fig.~\ref{fig:z11_1D},
the marginalized probability are non-Gaussian and distorted by their
correlations. This is particularly visible in the two-dimensional
likelihood in the plane $(\tau,\OmegaB h^2)$ (see
Fig.~\ref{fig:z11_2D}). Compared to the dark ages, varying $\tau$
affects the background signal amplitude at EoR through $\xH$ and
$\Tspin$. Compared to previous works derived with Fisher matrix
methods, our forecasts are slightly less optimistic, but as it should
be clear from Fig.~\ref{fig:z11_2D}, not including $\tau$ reduces the
expected error bars on various other cosmological parameters (see
below). The need of breaking degeneracies clearly suggests the
inclusion of additional data, either 21-cm from other redshift slices,
or CMB data.

We have also performed a Fisher matrix analysis under the same model
and experimental hypothesis, both on the convolved and nonconvolved
21-cm three-dimensional power spectra. Our results are reported in
Table~\ref{tab:z=11}. Without convolution, the strong degeneracy
between $\tau$ and $\ln \As$ leads to a ill-conditioned Fisher matrix
that can be hardly inverted. This can therefore induce important
numerical errors in the forecasts. Physically, this means that one
combination of parameters is completely unconstrained from the Fisher
point of view: the Gaussian approximation would predict some flat
directions for the likelihood. In such a situation, Fisher results can
still be derived provided the problematic parameters are artificially
removed to recover a well-behaved matrix.

The convolution with the frequency window function reduces
the degeneracy between the parameters $\tau$ and $\ln \As$ but still
does not cure the existence of flat directions. In
Table~\ref{tab:z=11}, we have reported the forecasts when removing
either $\ln \As$ or $\tau$ from the Fisher analysis. Forecast on the
other parameters can be slightly overpessimistic compared to the
MCMC. This is expected because, even with the convolution included,
the Fisher results still approximate any degeneracies as
linear. Figure~\ref{fig:z11_2D} shows that this is clearly not the case,
as for instance in the plane $(\OmegaB h^2,\tau)$ where the two-sigma
contour is actually multivalued. Let us notice that removing $\tau$
in the Fisher matrix yields an order of magnitude wrong prediction for
$\ln \As$, whereas removing $\ln \As$ gives a standard deviation for
$\tau$ compatible with the MCMC result (see
Table~\ref{tab:z=11}). These observations reinforce the need of MCMC
methods as soon as some degeneracies are not under control.

\begin{table}[h]
\begin{tabular}{|c|c|c|c|}
  \hline
  & $\OmegaB h^2$ & $\OmegaDM h^2$ & $\nS$ \\
  \hline
  M  & $9.5 \times 10^{-4} $ & $1.8 \times 10^{-3} $& $1.8 \times 10^{-3} $ \\
  F (no conv.) & $ 7.9 \times 10^{-4} $ & $1.5 \times 10^{-3} $ &$ 1.3 \times 10^{-3} $\\
  F (conv.)  & $9.2 \times 10^{-4} $ & $1.9 \times 10^{-3} $  & $ 1.7 \times 10^{-3} $ \\
  F (no conv.) & $ 8.0 \times 10^{-4} $ & $1.6 \times 10^{-3} $ &$ 1.3 \times 10^{-3} $\\
  F (conv.)  & $9.8 \times 10^{-4} $ & $2.0 \times 10^{-3} $  & $ 1.8 \times 10^{-3} $ \\
  \hline
  &  $ \tau $ &  $\ln(10^{10}\As)$  & $H_0$ \\
  \hline
  M  &$ 3.7 \times 10^{-3}$ & $3.7 \times 10^{-2}$ & 1.1 \\
  F (no conv.) & $3.0 \times 10^{-3}$ & - & $0.9$ \\
  F (conv.)  & $4.0 \times 10^{-3}$ & - & $1.0$  \\
  F (no conv.) & - & $0.24$ & $0.9$ \\
  F (no conv.) & - & $0.21$ & $1.0$ \\
  \hline     
\end{tabular}

\caption{Standard deviations on cosmological parameters for FFTT-only
  single redshift measurements at $z=11$ from both MCMC method (M) and
  Fisher matrix analysis (F). The Fisher results are given for both
  the nonconvolved and convolved power spectrum and by removing
  from the analysis one of the strongly degenerated parameters,
  either $\tau$ or $\ln(10^{10}\As)$.}
\label{tab:z=11}
\end{table}

\subsubsection{Multiredshifts tomography}
\label{sec:multi-redsh-tomogr}

Still with vanishing ionizing power spectra and a reionization model
having only $\tau$ varying, we can reduce the FFTT degeneracies by
adding extra redshift slices. As before, we have performed MCMC
analysis on the full sky mock data and Fisher matrix methods on the
associated 21-cm power spectrum.

\begin{figure}
\begin{center}
  \includegraphics[width=\columnwidth]{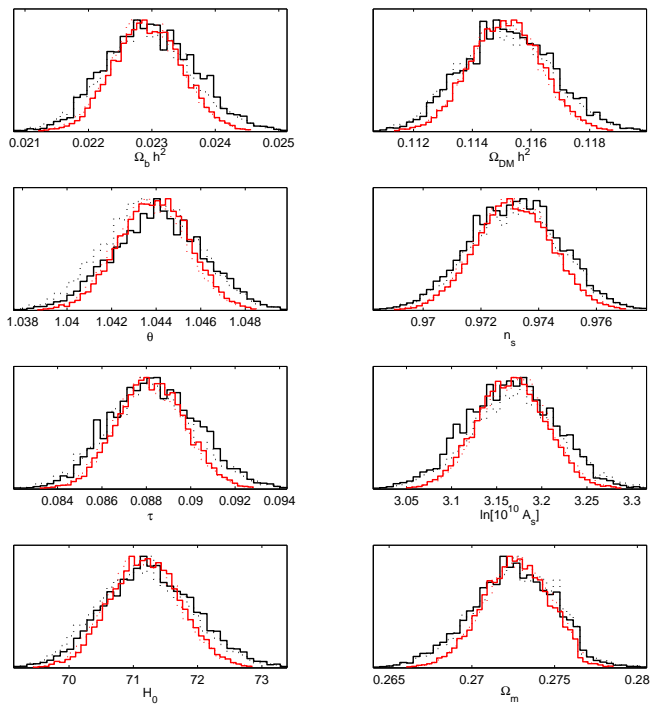}
  \caption{Marginalized posteriors (solid) and mean likelihood
    (dotted) from two and four observation redshifts at
    reionization. The black outer curves are for two redshift slices
    $z=11$, $z=12$ while the red inner ones for four redshift $z=10$,
    $z=11$, $z=12$ and $z=13$ (see also Fig.~\ref{fig:eor4z_2D}).}
  \label{fig:eor4z_1D}
\end{center}
\end{figure}

\begin{figure}
\begin{center}
  \includegraphics[width=\columnwidth]{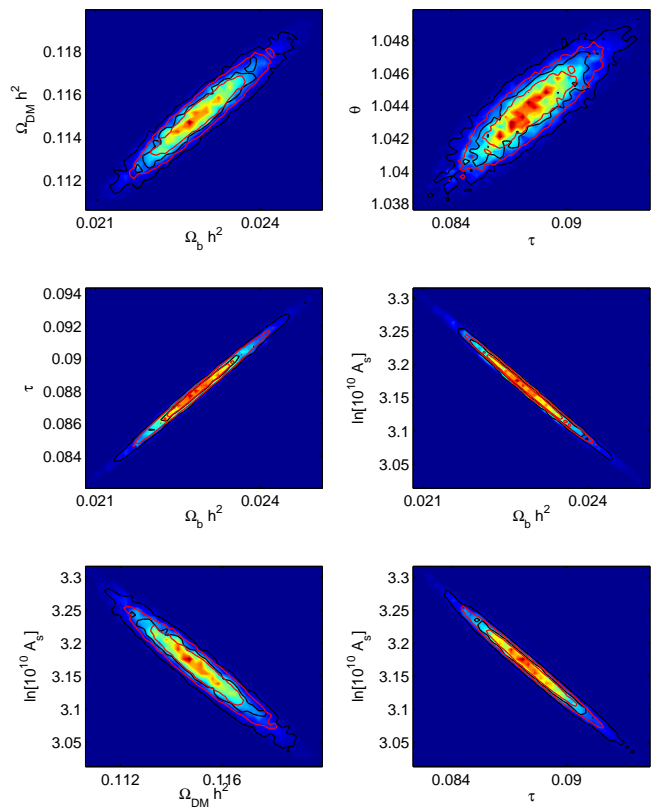}
  \caption{Two-dimensional posteriors obtained by combining the FFTT
    data from multiple redshift slices. The shading traces the mean
    likelihood associated with the combination of two redshifts $z=11$
    and $z=12$. The black largest contours are the associated one- and
    two-sigma confidence intervals whereas the smaller gray contours
    are those associated with combined data for four redshifts
    ($z=10$, $z=11$, $z=12$ and $z=13$). The degeneracies induced by
    $\tau$ are accordingly reduced.}
  \label{fig:eor4z_2D}
\end{center}
\end{figure}

In Figs.~\ref{fig:eor4z_1D} and \ref{fig:eor4z_2D} we have
represented the MCMC posteriors obtained by considering two redshift
slices, $z=11$, $z=12$, and four redshift slices $z=10$, $z=11$,
$z=12$, $z=13$. The telescope is, as before, the $10\,\km$ FFTT with
$\bwidth=10\,\MHz$ and all other parameters are the same.

Compared to $z=11$ alone, the degeneracies have been reduced. In fact,
one can check that a data analysis performed at $z=12$ alone would
produce similar posteriors as those obtained for $z=11$ with, however,
slightly different degeneracy directions in the parameter
space. Combining both can be viewed as keeping only the intercepting
probability contours of each redshift slice. For $\tau$, the standard
deviation ends up being reduced by almost a factor of two, while this
is not as significant for the other parameters. Their posteriors do no
longer exhibit ``shoulders'' or evident non-Gaussian shapes, however
Fig.~\ref{fig:eor4z_2D} shows that they still are correlated along
some peculiar directions. Adding more redshifts again improves the
situation, but the correlation trend does not disappear due to the
common physical origin of all 21-cm signals at reionization. The inner
red curves in Fig~\ref{fig:eor4z_1D} and Fig~\ref{fig:eor4z_2D} are
the same analysis performed on four redshift slices, scanning the
whole reionization duration: $z=10$, $z=11$, $z=12$ and $z=13$. The
standard deviations are slightly reduced, but less than the expected
$\sqrt{2}$ factor because both $z=10$ and $z=13$ correspond to a low
emission signal (see Fig.~\ref{fig:sncls}). Let us notice that a
bandwidth $\bwidth=10\,\MHz$ corresponds in term of redshift to a
width $\delta z\simeq 1$ (at $z=11$). Requiring more redshift slices
at reionization would therefore also demands a smaller bandwidth.

Let us turn now to the Fisher analysis. The forecasts for the four
redshifts $z=10$, $z=11$, $z=12$ and $z=13$ and the convolved 21-cm
power spectrum are given in Table~\ref{tab:cmbeor4z}. Our results agree
relatively well with those from the MCMC method, even if the forecasts
are found to be about $20\%$ more stringent with the Fisher matrix
analysis. We have also checked that the degeneracy directions between
parameters were matching the ones obtained from the MCMC and
represented in Fig.~\ref{fig:eor4z_2D}.
\subsubsection{Adding Planck-like CMB data}

\label{sec:addcmb}

\begin{figure}
\begin{center}
  \includegraphics[width=\columnwidth]{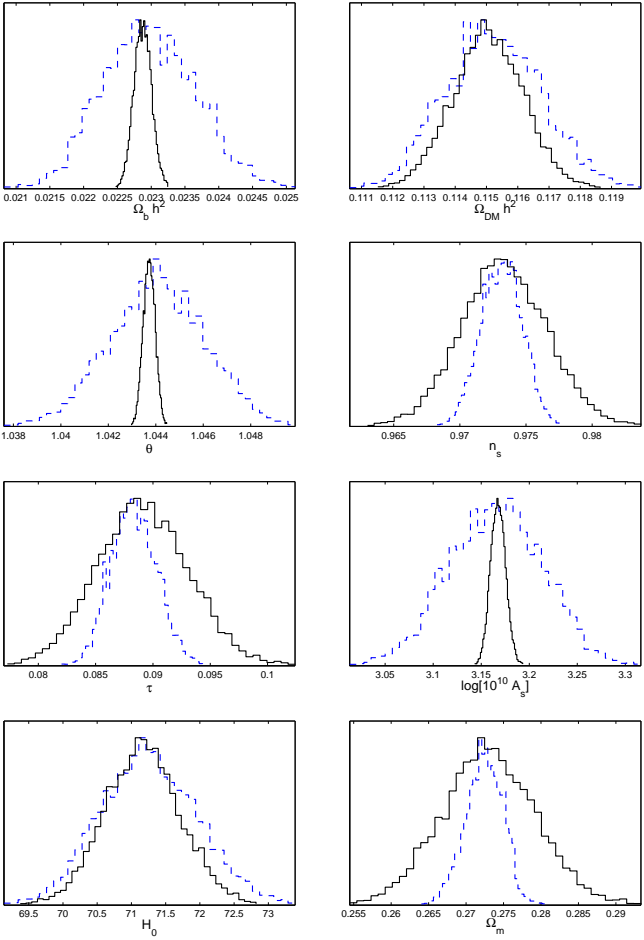}
  \caption{Typical marginalized posterior distributions for a
    Planck-like CMB experiment (black solid lines). The dashed blue
    curves are the two redshift forecasts for FFTT (same as in
    Fig.~\ref{fig:eor4z_1D}).}
  \label{fig:planck}
\end{center}
\end{figure}

The previous results call for the incorporation of other cosmological
data having different correlations. In the following, we incorporate
CMB data and consider a typical Planck-like experiment. For our
purpose, we have generated mock CMB data exactly as described in
Sec.~\ref{sec:fftt}, but for a full sky CMB experiment ($\fsky=1$),
with a Gaussian beam of resolution $\fwhm=7'$ and a constant noise for
each pixel given by $\mean{\delta \Tn^2} = 2 \times 10^{-4}\,\muK^2$
(twice for the $E$-mode polarization)~\cite{2010AA520A9L,
  2010AA520A11A}. Performing a MCMC analysis with Planck-like mock
data alone, and for the same fiducial parameter values, yields the
posterior plotted in Fig.~\ref{fig:planck}. To ease the comparison, we
have reported on the same figure the two redshifts' FFTT forecasts.

CMB data give much stronger constraints on $\OmegaB h^2$, $\As$ and
$\theta$ than the considered $10\,\km$ FFTT design. This is expected
because the overall amplitude of the CMB signal depends mainly on
$\As$ and not $\OmegaB h^2$ contrary to the 21-cm
brightness. Concerning $\theta$, it is an optimal parameter for CMB by
design whereas $H_0$ shows similar posteriors for both CMB and 21
cm. Interestingly, the spectral index is, by a factor of two, more
constrained by 21-cm. One can also check in Fig.~\ref{fig:z11_1D} that
this would also the case with only one observation redshift at
$z=11$. This is again due to the huge level arm provided by the FFTT
resolution on the power spectrum shape. Moreover,
Fig.~\ref{fig:planck} shows that, in spite of the degeneracies, the
marginalized posterior for $\tau$ remains sharper for FFTT than for
CMB. This could wrongly suggest that adding CMB data may not help to
improve the FFTT constraints over $\tau$. However, $\tau$ is mostly
correlated with $\OmegaB h^2$ which is well bounded by CMB alone. As a
result, CMB data are expected to indirectly improve the FFTT
accuracy on $\tau$.

\begin{figure}
\begin{center}
  \includegraphics[width=\columnwidth]{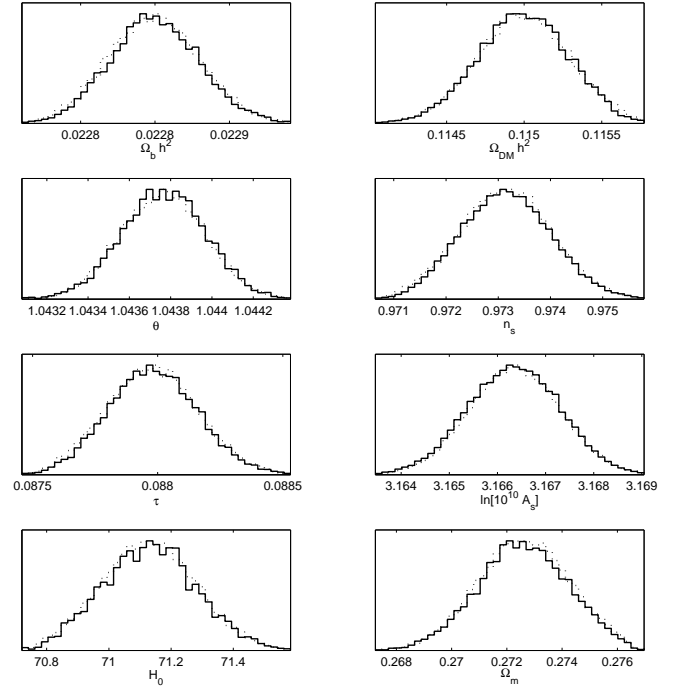}
  \caption{Marginalized posteriors from the combination of the FFTT
    four redshifts data ($z=10$, $z=11$, $z=12$ and $z=13$) with
    Planck-like CMB data. The reduced variance on $\OmegaB h^2$ coming
    from CMB data kills most of the degeneracies associated with the
    21-cm signal. Four redshifts' FFTT without CMB is represented in
    Fig.~\ref{fig:eor4z_1D}.}
  \label{fig:cmbeor4z_1D}
\end{center}
\end{figure}

\begin{figure}
\begin{center}
  \includegraphics[width=\columnwidth]{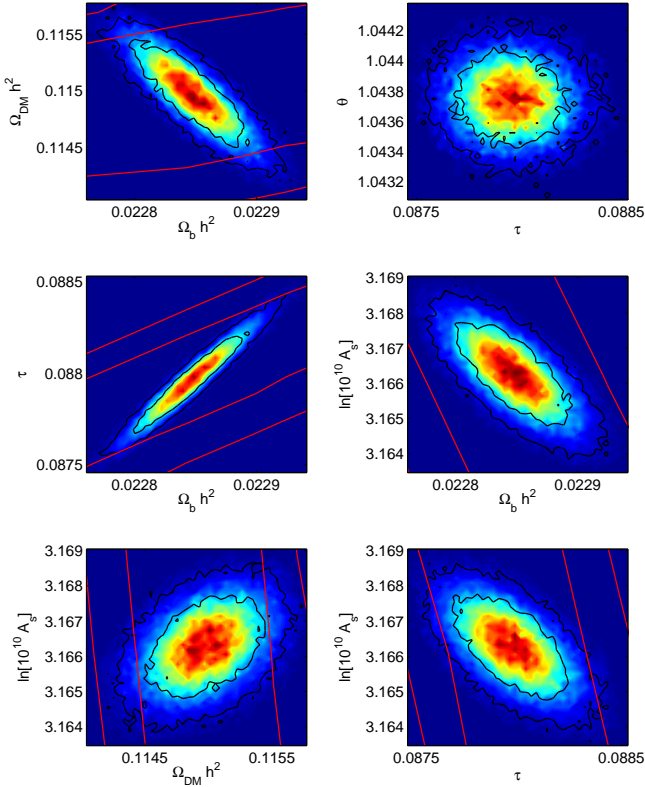}
  \caption{Two-dimensional marginalized probability distribution for
    the combined four redshifts' FFTT and Planck-like CMB data. The
    inner black solid lines are the one- and two-sigma confidence
    intervals, the shading traces the mean likelihood. The outer red
    lines are the one- and two-sigma confidence intervals for four
    redshifts' FFTT data alone. The precise determination of $\OmegaB
    h^2$ from CMB strongly reduces the $(\tau,\OmegaB h^2)$
    correlations associated with the 21-cm signal.}
  \label{fig:cmbeor4z_2D}
\end{center}
\end{figure}

\begin{table*}
\begin{tabular}{|c|c|c|c|c|c|c|c|c|c|}
\hline
Data & Method &$\OmegaB h^2$ & $\OmegaDM h^2$ & $\theta$ & $\nS$ & $\tau$
& $\ln(10^{10}\As)$ & $H_0$  \\
\hline
FFTT alone & M & $5.4\times 10^{-4}$ &  $1.2 \times 10^{-3}$   & $1.6
\times 10^{-3}$  &  $1.3 \times 10^{-3}$   &  $1.5 \times 10^{-3}$  &
$3.8 \times 10^{-2}$  & $0.55$   \\
 & F & $4.2 \times 10^{-4} $& $9.5 \times 10^{-4} $& $1.3 \times 10^{-3}$& $1.0 \times 10^{-3}$ & $1.2 \times 10^{-3} $
 &$2.9 \times 10^{-2} $ & $ 0.45$  \\
 & F ($B=0.1\, \MHz$, no conv.) & $2.3 \times 10^{-5} $ & $1.6 \times 10^{-4}$ &$1.4 \times 10^{-4}$ &$1.7 \times 10^{-4}$ & $4.9 \times 10^{-5}$ & $ 2.3 \times 10^{-3}$ & $4.9 \times 10^{-2}$ \\ 
\hline
FFTT + CMB & M & $3.0 \times 10^{-5}$ & $2.9 \times 10^{-4}$ & $2.0 \times
10^{-4}$ & $8.6 \times 10^{-4}$  & $1.8\times 10^{-4}$ & $9.5 \times
10^{-4}$ & $0.15$  \\
 & F & $ 2.8 \times 10^{-5} $  & $2.6 \times 10^{-4}$ & $2.2 \times 10^{-4} $  & $7.2 \times 10^{-4}$ & 
 $1.6 \times 10^{-4}$& $9.0 \times 10^{-4}$& $0.13$ \\
 
& F ($B=0.1\, \MHz$, no conv.) & $8.1 \times 10^{-6} $ & $ 3.9 \times 10^{-5}$ & $8.3 \times 10^{-5} $   & $5.7 \times 10^{-5}$ &$3.6 \times 10^{-5}$ & $ 3.4 \times 10^{-4}$  & $1.2 \times 10^{-2}$  \\
\hline
\end{tabular}
\caption{Expected standard deviations for the basic parameters
  in the case of four redshifts' FFTT alone and complemented with
  Planck-like CMB data. The reionization model has only one varying
  parameter $\tau$, the ionizing power spectra are assumed to be
  negligible (for the MCMC method, $H_0$ is a derived parameter
  and is reported in the last column). The method refers to either
  MCMC (M) full sky analysis or the Fisher matrix (F) on the 21cm
  three-dimensional power spectrum (with convolution). The third and sixth lines show
  the Fisher results (without convolution) in the small bandwidth
  limit $B=0.1 \,\MHz$, with $\delta z=0.5$, for which $40$ $\upara$ modes would be accessible.}
\label{tab:cmbeor4z}
\end{table*}

We have checked that this is indeed the case by performing a MCMC
analysis based on Planck-like CMB data together with the four
redshifts' FFTT data. The resulting posterior probability distributions
are represented in Fig~\ref{fig:cmbeor4z_1D}. The expected standard
deviation on $\tau$ is reduced by an order of magnitude. In
Fig.~\ref{fig:cmbeor4z_2D}, the two-dimensional posterior confirms
that such an improvement comes from the significant reduction in the
variance of $\OmegaB h^2$ due to CMB data. For comparison, we have
also reported in this figure the one- and two-sigma confidence
intervals for the four redshifts' FFTT alone (see
Fig.~\ref{fig:eor4z_2D}). For some parameters, the improvement is such
that they may lie outside the displayed plane.

The forecasts derived from the Fisher matrix are reported in
Table~\ref{tab:cmbeor4z}. As the degeneracies are even more reduced by
the inclusion of CMB data, the Fisher results agree well with those
from the MCMC analysis. The degeneracy axes associated with
one-sigma ellipses are also found to be in agreement with the MCMC
two-dimensional posteriors (not represented).  

As the Fisher matrix approach can be trusted in this regime, one can
discuss the effect of increasing the number of modes by reducing the
bandwidth. For $B=0.1 \, \MHz$, still using the same four mean
redshifts, with a bin size of $\delta z=0.5$, it is no longer
necessary to include the convolution. In that case, there are about
$40$ parallel modes accessible along the line-of-sight, up to a
nonlinear cutoff fixed at $\kpara = 2\, \Mpc^{-1}$~\cite{Mao:2008ug}
(see Table~\ref{tab:cmbeor4z}).

In the following sections, we relax some of the reionization
assumptions and consider that duration and maximal spin temperature
are no longer known.

\subsubsection{Spin temperature evolution and reionization duration}

The reionization model of the previous sections involved a
unique varying parameter $\tau$. One may now wonder how the forecasts
are affected by the reionization duration $\Deltaz$ and the maximal
spin temperature $\TspinMax$. Again, the fiducial model is unchanged,
i.e., has $\fidDeltaz=0.5$ and $\fidTspinMax=10000\,\K$.

\begin{figure}
\begin{center}
  \includegraphics[width=\columnwidth]{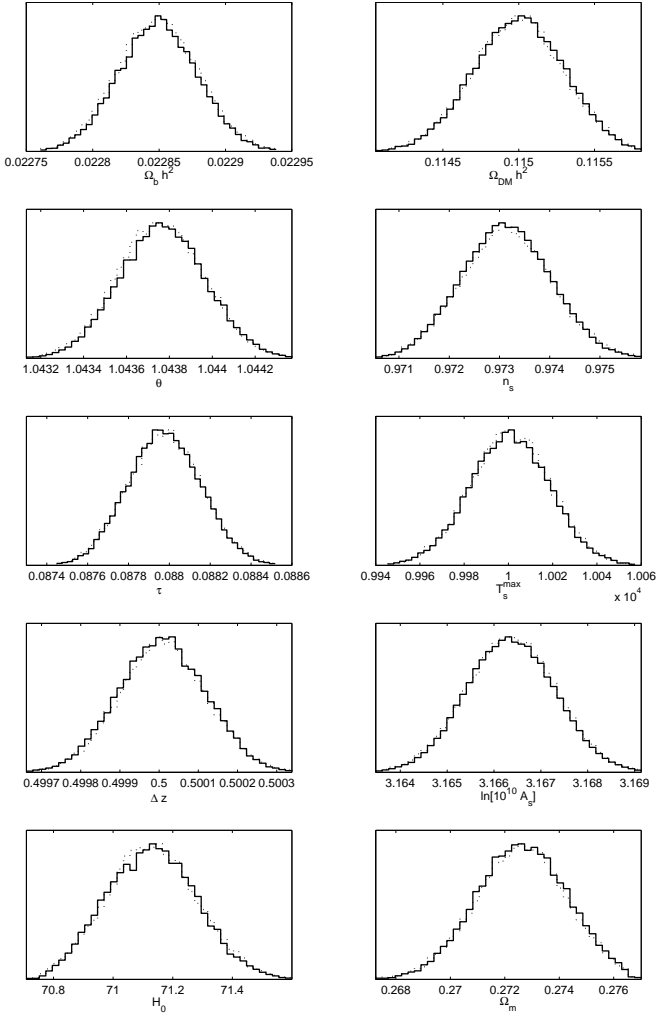}
  \caption{Posterior probability distributions from combined FFTT four
    redshifts and Planck-like CMB data. The background reionization
    model has three varying parameters, $\tau$, $\Deltaz$ and
    $\TspinMax$, all well reconstructed.}
  \label{fig:dztscmbeor4z_1D}
\end{center}
\end{figure}

\begin{figure}
\begin{center}
  \includegraphics[width=\columnwidth,height=1.5\columnwidth]{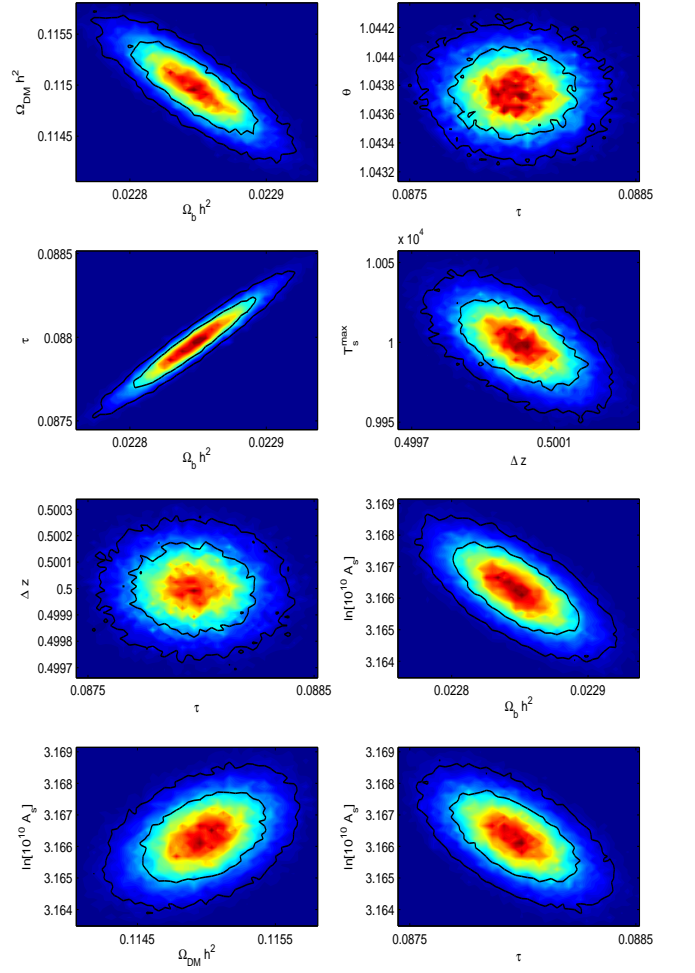}
  \caption{Two-dimensional posterior probability distributions
    associated with the combined FFTT four redshifts and Planck-like CMB
    data (see also Fig.~\ref{fig:dztscmbeor4z_1D}). The background
    reionization model parameters, $\tau$, $\Deltaz$ and $\TspinMax$
    are weakly correlated and well reconstructed.}
  \label{fig:dztscmbeor4z_2D}
\end{center}
\end{figure}

Figures~\ref{fig:dztscmbeor4z_1D} and \ref{fig:dztscmbeor4z_2D} show
the probability distributions obtained from the MCMC exploration based
on the same data, i.e. four redshifts' FFTT plus CMB Planck-like
data. The two additional reionization parameters are very well
constrained. Even for the large chosen fiducial value $\fidTspinMax$,
the expected standard deviation does not exceed $20\,\K$. This directly
comes from the FFTT tomography. Along the four redshifts probed, the
21cm signal is very sensitive to the neutral hydrogen fraction
$\xH(z)$ and the value of $\Tspin(z)$ thereby providing accurate
forecasts for the related parameters. As visible in
Fig.~\ref{fig:dztscmbeor4z_2D}, $\TspinMax$ is slightly correlated
with $\Deltaz$ but this is expected in view of our smooth reionization
model (see Fig.~\ref{fig:btbg}). Increasing $\TspinMax$ at a given
redshift mimics a reduction in $\Deltaz$. The correlations remain
under control because we use more than one redshift. These results
also suggest that taking the asymptotic limit $\Tspin \rightarrow
\infty$ at reionization is not always justified and throws away some
potentially available information.

\begin{table}
\begin{tabular}{|c|c|c|c|c|}
  \hline
  &$\OmegaB h^2$ & $\OmegaDM h^2$  & $\nS$ & $H_0$ \\
  \hline
  M & $3.0 \times 10^{-5}$ & $2.9 \times 10^{-4}$ & $8.8 \times 10^{-4}$  & $0.15$  \\
  F  & $2.9 \times 10^{-4} $ & $2.6   \times 10^{-4}$ &  $7.4 \times 10^{-4}$ & $ 0.14$ \\
  \hline
  &$\tau $ & $\ln(10^{10}\As)$ & $\TspinMax (\K)$ & $\Deltaz$ \\
  \hline
  M & $1.8 \times 10^{-4}$ & $9.5 \times 10^{-4}$ & $18$ & $1.1 \times 10^{-4}$  \\
    F  & $ 1.6 \times 10^{-4} $ & $9.4 \times 10^{-4} $ & - & $2.7 \times 10^{-4}$ \\     
  \hline     
\end{tabular}
\caption{Standard deviations for the combined four redshifts' FFTT and
  CMB for a reionization model with three varying parameters $\tau$,
  $\Deltaz$ and $\TspinMax$. There is no loss of accuracy compared to
  Table.~\ref{tab:cmbeor4z} showing that redshift tomography allows
  the complete reconstruction of the reionization parameters. The
  labels ``M'' and ``F'' respectively refer to MCMC and Fisher matrix
  analysis (with convolution). The Fisher matrix has to be regularized
  at the expense of losing information by removing one of the
  degenerated parameters, here $\TspinMax$.}
\label{tab:dztscmbeor4z}
\end{table}

We find that the Fisher matrix analysis, with convolution included,
does not resolve completely the degeneracy between $\TspinMax$ and
$\Deltaz$ thereby rendering the Fisher matrix singular. As before, one
needs to remove the incriminated flat directions, and we had to pull
out $\TspinMax$ from the analysis\footnote{Removing $\Deltaz$ still
  yields an almost singular matrix generating spurious numerical
  errors at inversion.}. In this case, the predicted standard
deviations are in good agreement with the MCMC results, whereas larger
differences between the two methods appear for $\Deltaz$. This is again
not surprising because those parameters are correlated (see
Fig.~\ref{fig:dztscmbeor4z_2D}).

In the next section, we relax the assumption that the ionizing power
spectra are negligible and consider their impact on the model
parameter forecasts.

\subsubsection{Nuisance ionization power spectra}

\begin{figure*}
\begin{center}
  \includegraphics[width=\columnwidth]{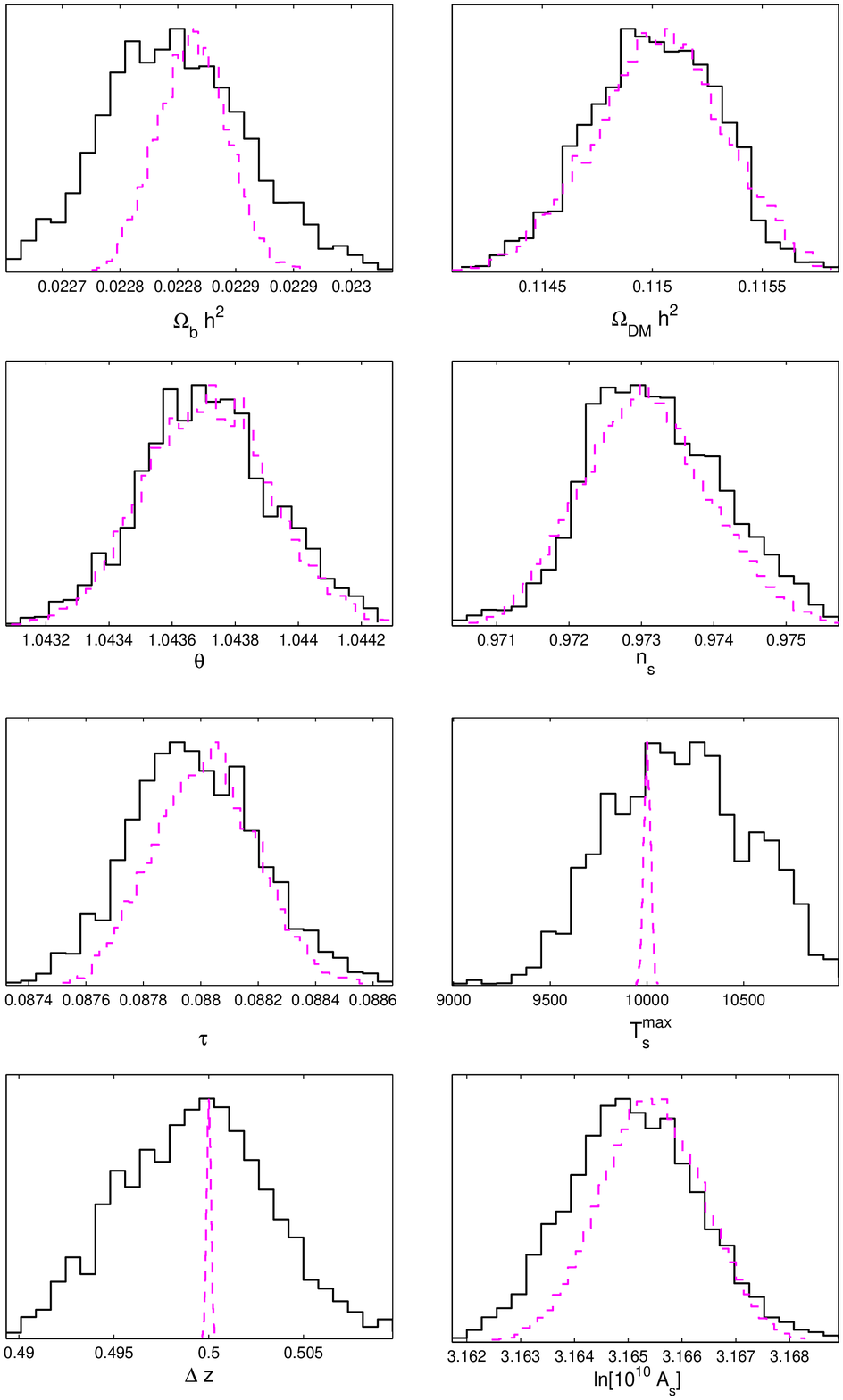}
  \includegraphics[width=\columnwidth]{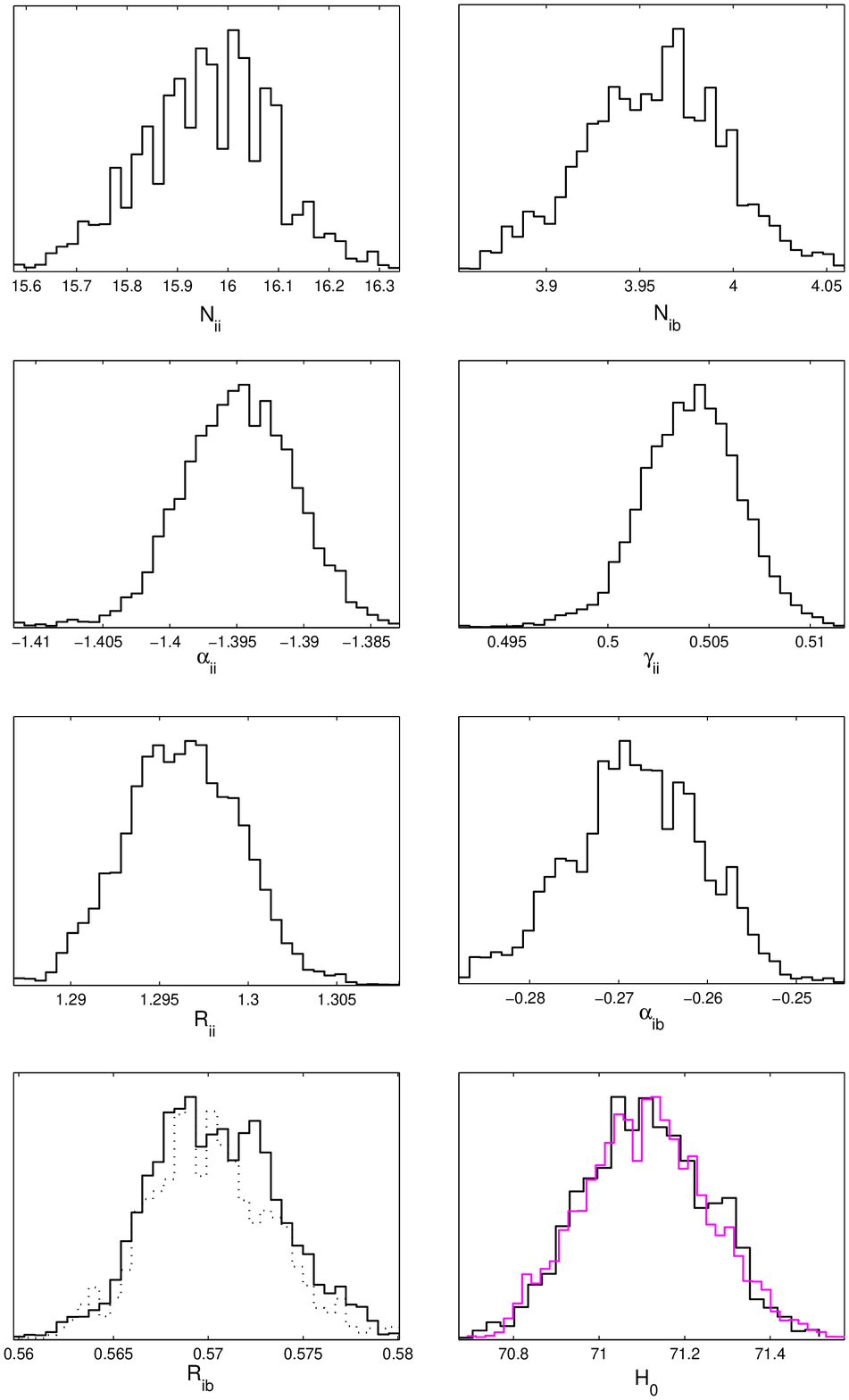}
  \caption{Posterior probability distributions in presence of the
    ionizing power spectra. The background reionization model has
    three varying parameters, $\tau$, $\Deltaz$ and $\TspinMax$ while
    the ionizing spectra involves seven nuisance parameters $\Nii$,
    $\alphaii$, $\gammaii$, $\Rii$, $\Nib$, $\alphaib$ and $\Rib$. The
    posteriors of Fig.~\ref{fig:dztscmbeor4z_1D} have been reported as
    dashed curves to ease comparison.}
  \label{fig:idztscmbeor4z_1D}
\end{center}
\end{figure*}

As in Ref.~\cite{Mao:2008ug}, we consider a less ideal situation in
which the ionizing sources affect the 21-cm perturbations by switching
on the extra power spectra $\Pib$, $\Pii$ of
Eq.~(\ref{eq:PTb}). According to the discussion of
Sec.~\ref{sec:fore}, the redshift evolution of our toy ionizing
spectra is completely encoded into $\xH(z)$ and $\xI(z)$; as in
Eq.~(\ref{eq:P024}). In total, this adds seven nuisance parameters to
the analysis, namely $\Nii$, $\Rii$, $\alphaii$, $\gammaii$, $\Nib$,
$\Rib$ and $\alphaib$. In a more realistic situation, the ionizing
spectra may involve more parameters and have certainly more
complicated shapes, but their redshift evolution should still be
correlated with the background reionization history.

\begin{figure}
\begin{center}
  \includegraphics[width=\columnwidth]{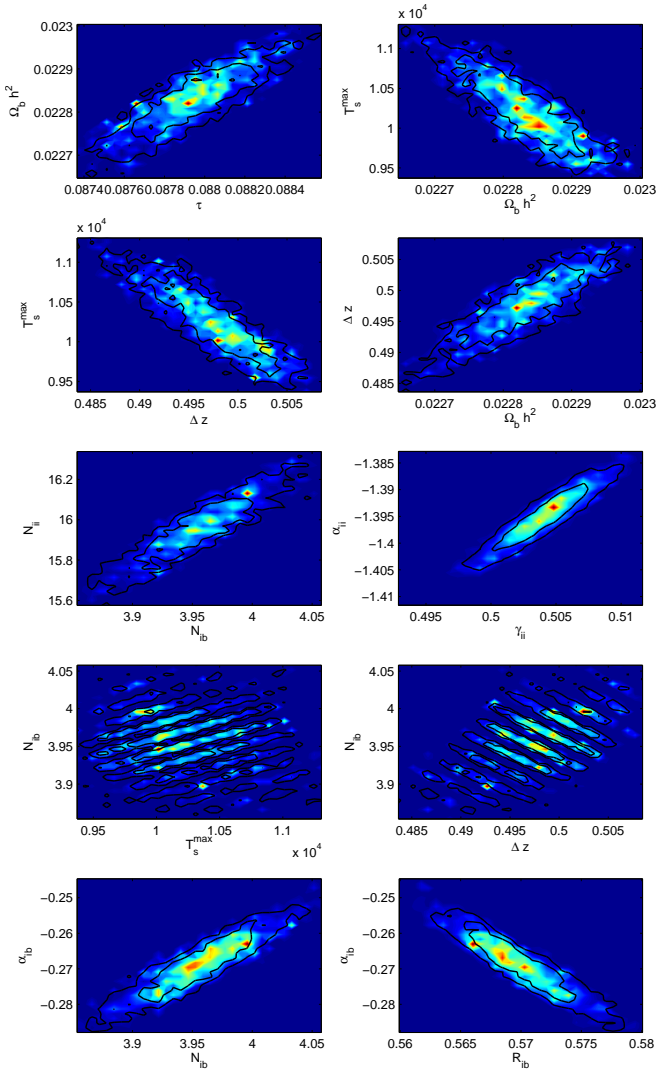}
  \caption{Two-dimensional posterior probability distributions in
    presence of the ionizing power spectra. The correlations induced
    by the nuisance parameter $\Nib$ severely damages the expected
    constraints on the background reionization parameters $\Deltaz$
    and $\TspinMax$. The multivalued confidence intervals are the
    results of the acoustic oscillations (see text).}
  \label{fig:idztscmbeor4z_2D}
\end{center}
\end{figure}

Using again the four redshifts' FFTT plus Planck-like CMB data, we have
plotted in Fig.~\ref{fig:idztscmbeor4z_1D} and
\ref{fig:idztscmbeor4z_2D} the resulting marginalized probability
distributions when the ionizing spectra are switched on. The fiducial
values of all parameters are the same as before while the seven
nuisance parameters have been fixed to the fiducial values of
Table~\ref{tab:ioparams}.

The most important result is that the posteriors for the cosmological
parameters are almost unchanged compared to
Fig.~\ref{fig:dztscmbeor4z_1D} thereby confirming that the presence of
nuisance ionizing spectra should not prevent cosmological parameter
estimation~\cite{Mao:2008ug}. The expected standard deviations have
been reported in Table~\ref{tab:idztscmbeor4z}, and up to an increase
for $\OmegaB h^2$ and $\tau$, the bounds of all the other cosmological
parameters are identical to the ones of
Table.~\ref{tab:dztscmbeor4z}. Let us stress again that this result
incorporates marginalization over all background reionization
histories accessible within our model.

The situation is different for the background reionization
parameters. The standard deviation associated with the posteriors of
$\Deltaz$ and $\TspinMax$ is more than one order of magnitude larger
than those obtained without ionizing spectra. As can be seen in
Fig.~\ref{fig:idztscmbeor4z_2D}, the two-dimensional posteriors
exhibit strong correlations between the nuisance parameters and
$\Deltaz$ and $\TspinMax$. This can be understood from
Fig.~\ref{fig:iopower}. Around the redshift of maximum sensitivity
($z\simeq11$), the total ionizing spectra is dominated by $\Pib$. Its
amplitude $\Nib$ therefore affects the overall observable signal and
should therefore correlate to all the other amplitude related
parameters. However, since we use CMB data, the parameters $\OmegaB
h^2$ and $\tau$ are relatively well constrained, independently of the
values of $\Nib$. They are therefore weakly affected by these new
correlations. It remains the neutral fraction $\xH$, which is
precisely given by the background reionization history, i.e. by the
value of $\Deltaz$ and $\TspinMax$. One can see in
Fig.~\ref{fig:idztscmbeor4z_2D} that the two-dimensional posteriors in
the planes $(\Deltaz,\Nib)$ and $(\TspinMax,\Nib)$ are indeed
degenerated and even multivalued. The ``strips'' of maximum posterior
values are due to the presence of the acoustic oscillations which
favour only some peculiar combination of ionizing amplitude and
neutral fraction.

\begin{table}
\begin{tabular}{|c|c|c|c|c|c|}
  \hline
  &$\OmegaB h^2$ & $\OmegaDM h^2$ &  $H_0$    &  $\Nib$ & $\Nii$ \\
  \hline
  M & $5.8 \times 10^{-5}$ & $3.0 \times 10^{-4}$ & $0.15$ & $0.04$ & $0.13$ \\
  F & $ 8.1 \times 10^{-5} $ &$4.1 \times 10^{-4} $ &$0.16$& $0.03$&$ 0.10$  \\
  \hline
  &$\TspinMax (\K)$ & $\Deltaz$ & $\ln(10^{10}\As)$ & $\Rib $ & $ \Rii$  \\
  \hline
  M & $370$ & $4.2 \times 10^{-3}$ & $1.2 \times 10^{-3}$ & $3.4 \times 10^{-3}$
  & $3.3 \times 10^{-3}$ \\
  F & - & $9.4 \times 10^{-4}$ & $1.6\times 10^{-3}$ & $ 3.6 \times 10^{-3}$& $3.0 \times 10^{-3}$  \\
  \hline     
  &  $\nS$  & $\tau$ &  $\gammaii $ & $\alphaib $ & $\alphaii$ \\
  \hline
  M & $9.8 \times 10^{-4}$   & $2.0\times 10^{-4}$   & $2.5\times
  10^{-3}$  & $7.5\times 10^{-3}$ & $4.3\times 10^{-3}$  \\
  F &  $ 1.7 \times 10^{-3} $ &  $ 3.2 \times 10^{-4} $  & $2.5 \times
  10^{-3}$ & $1.1 \times 10^{-2}$
  & $ 4.3 \times 10^{-3}$ \\
  \hline
\end{tabular}
\caption{Standard deviations in presence of the ionizing power
  spectra. The mock data are, as before, four redshifts' FFTT and
  Planck-like CMB. Although there is no loss of accuracy compared to
  Table.~\ref{tab:dztscmbeor4z} for the cosmological parameters, the
  expected constraints on the background reionization parameter $\Deltaz$
  and $\TspinMax$ are severely damaged due to the appearance of new
  degeneracies (see Fig.~\ref{fig:idztscmbeor4z_2D}).}
\label{tab:idztscmbeor4z}
\end{table}

When marginalizing over the cosmological and background reionization
parameters, Fig.~\ref{fig:idztscmbeor4z_1D} and
Table~\ref{tab:idztscmbeor4z} show that the nuisance parameters $\Nib$,
$\alphaib$ and $\Rib$ can actually be inferred. This is not surprising
as they encode the dominant contribution for the ionizing power
spectra. On the other hand, although the posteriors of $\Nii$, $\Rii$,
$\alphaii$ and $\gammaii$ peak at some preferred values, these are
significantly biased compared to the fiducial ones. Again, this can be
understood from Fig.~\ref{fig:iopower}. The $\Pii$ contribution only
shows up at small scales, typically $\ell \gtrsim 10000$, where the
noise power spectrum becomes of comparable amplitude to the signal
(see Fig.~\ref{fig:sncls}).

Unsurprisingly, the above-mentioned degeneracies are problematic for a
Fisher matrix analysis. As before, the only solution to render the
Fisher matrix regular was to remove the parameter $\TspinMax$, whereas
removing $\Deltaz$ would not fix the ill conditioning. As can be seen
in Table~\ref{tab:idztscmbeor4z}, the Fisher method (convolution
included) has difficulty to reproduce all of the MCMC results
accurately. For some parameters, precisely the ones which remains
weakly sensitive to the ionization parameters, the Fisher forecasts
match well with the MCMC. This is the case for the standard
cosmological parameters (up to $\OmegaB h^2$). For the background
reionization parameters, as for instance $\Deltaz$, the Fisher
expected variances can be out by a factor five, which slightly bias
the expected variance of $\tau$ and $\OmegaB h^2$. Turning off
$\TspinMax$ also kills the multimodal behaviour of the likelihood and
the Fisher results end up luckily giving the correct results for the
ionizing power spectra amplitudes, $\Nii$ and $\Nib$.

To summarize, we have found that even in presence of ionizing power
spectra, combining multiredshift FFTT data with CMB data can still be
used to reconstruct the background reionization history together with
the more usual set of cosmological parameters. The expected variances
of the reionization parameters are nevertheless severely damaged
compared to no ionizing sources.

\section{Conclusion}

In this paper, we have quantitatively shown that omniscopes could be
used to constrain the background reionization history while being not
sensitive to the zero mode of the brightness temperature.

For this purpose, we have considered a simple but consistent
reionization model completely determined by the total optical depth
$\tau$, the reionization duration $\Deltaz$ and the asymptotic spin
temperature $\TspinMax$. We have used both Fisher matrix approaches on
the three-dimensional power spectrum and MCMC methods on full sky
simulated data to forecast the expected variance of all model
parameters. Our results suggest that it is crucial to combine multiple
redshifts at the EoR with CMB data to keep all degeneracies under
control, in particular owing to an accurate determination of $\OmegaB
h^2$. When this is not the case, we have shown that the Fisher
predictions can be quite inaccurate, eventually being overoptimistic,
but also overpessimistic when the conditions for the Cram\'er-Rao
inequalities are no longer met. For those situations, only the MCMC
methods ends up being usable.

Within combined data, the perspectives are quite good as the
background history can be fully reconstructed, even in the presence of
correlated nuisance ionizing sources. Although we have discussed only
one kind of FFTT telescope, our results could be easily generalized
to other configurations by a proper rescaling of the resolution and
noise parameters of Sec.~\ref{sec:models}. Still, our work would need
to be extended with a more complete reionization model, eventually
adjusted to numerical simulations in order to include a realistic
dependence between the reionization parameters and the redshift
evolution of the ionization power spectra. Another extension would be
the inclusion of exotic reionization sources, such as a decaying or
annihilating dark matter component.

\section{Acknowledgements}

It is a pleasure to thank Alex Hall for fruitful discussions and
comments. S.C. is supported by the Wiener Anspach foundation;
L.L.H. acknowledges partial support from the European Union FP7 ITN
INVISIBLES (Marie Curie Actions, Grant No. PITN-GA-2011-289442), the
Galileo Galilei Institute for Theoretical Physics and the INFN for
hospitality, the Belgian Federal Science Policy Office through the
Interuniversity Attraction Pole Grant No. P7/37 and the
‘‘FWO-Vlaanderen’’ for a postdoctoral fellowship (Grant No. 1271513N)
and for the Grant No. G.0114.10N. C.R. is partially supported by the
ESA Belgian Federal PRODEX Grant No.~4000103071 and the
Wallonia-Brussels Federation grant ARC No.~11/15-040. The work of
M.H.G.T. is partially supported by the grant ARC ``Beyond Einstein:
fundamental aspects of gravitational interactions''.

\appendix

\section{Boltzmann equations}
\label{sec:boltz}

In this appendix, we recap some of the equations driving the dynamics
of the 21-cm brightness fluctuations as they are derived in
Ref.~\cite{Lewis:2007kz}. Denoting by $f(\eta, \epsilon)$ the photon
distribution function of 21-cm photons, where $\epsilon$ is the
redshifted energy of 21-cm photons ($\Tb = \epsilon f/2$ in
  the Rayleigh-Jeans approximation), the multipole components of the
photon distribution function $F_\ell(\eta, \epsilon,k)$ are defined
by
\begin{equation}
\delta f(\eta,\epsilon,\n) = 4 \pi  \sum_{\ell,m} \int \dfrac{\ud
  \bk^3}{(2 \pi)^{3/2}} (-i)^{\ell} F_\ell \, Y_{\ell}^{m*}(\bk)
Y_{\ell}^m(\n),
\end{equation}
where the $Y_{\ell}^m$ are the spherical harmonics. From the perturbed
Boltzmann equation, they are found to verify~\cite{Lewis:2007kz}
\begin{widetext}
\begin{equation}
\begin{aligned}
  F_\ell(\eta_0,\epsilon,k) & = e^{-\tauC} \left\{ f_\epsilon \left[
    \Dmono + \Psi + \frac{\rscatE \partial_\eta\Phi}{\calH} +
    \left(\rscatE - 1 \right) \left(\DHI -\DTspin + \Psi \right)
    \right] - g_\epsilon \Psi \right \}_{\etaE} j_\ell(k \Delta\etaE)
  \\ & - e^{-\tauC} \left[\frac{\rscatE f_\epsilon}{\calH}
    \left(\partial_\eta v +\calH v - k \Psi \right) + g_\epsilon v +
    f_\epsilon \frac{\Trad}{\Tspin-\Trad} \left(v_\gamma-v\right)
    \right]_{\etaE} j_\ell'(k\Delta\etaE) +\left(\rscatE e^{-\tauC}
  f_\epsilon \Dv\right)_{\etaE} j_\ell''(k \Delta\etaE) \\ & -
  \int_{\etaE}^{\eta_0} \ud \eta \, (\partial_\eta\tauC) e^{-\tauC}
  \left\{ \left(F_0 - g_\epsilon \Psi \right) j_\ell(k \Delta\eta) -
  g_\epsilon v j_\ell'(k \Delta\eta) + \frac{F_2}{4} \left[ 3
    j_\ell''(k \Delta\eta) + j_\ell(k \Delta\eta) \right] \right\}
  \\ & - g_\epsilon \int_{\etaE}^{\eta_0} \ud \eta \, e^{-\tauC}
  \partial_\eta(\Phi + \Psi) j_\ell(k\Delta\eta) - e^{-\tauC} f_\epsilon
  \frac{\Trad}{\Tspin - \Trad} \sum_{\ell'=2}^\infty
  (2\ell'+1)\Theta_{\ell'} i^{\ell'}
  P_{\ell'}\left(-\frac{i}{k}\frac{\ud}{\ud \etaE}\right) j_\ell(k
  \Delta\etaE),
  \label{eq:fl21cm}
\end{aligned}
\end{equation}
\end{widetext}
In this equation, the conformal distance to the redshift of
observation is given by$\Delta \etaE\equiv\etaE-\eta_0$, $\tauC$ is
the Thomson scattering optical depth, $\tauE$ the optical depth of 21
cm photons and $\rscatE \equiv \tauE e^{-\tauE}/(1-e^{-\tauE})$
encodes the amount of rescattered 21-cm photons. The Legendre
polynomials have been denoted $P_\ell$ while $\Theta_\ell$ stand for the
multipole moments of the CMB temperature
anisotropies~\cite{Lewis:2007kz}. Both $f_\epsilon$ and $g_\epsilon$
are functions of $\epsilon$ only conveniently defined from the
background distribution function
\begin{equation}
f(\eta,\epsilon) \equiv f_\epsilon
\dfrac{1-e^{-\tausf(\eta)}}{1-e^{-\tauE}}\,,
\end{equation}
and its derivative
\begin{equation}
\epsilon \partial_\epsilon f(\eta, \epsilon) \equiv g_\epsilon
\dfrac{1-e^{-\tausf(\eta)}}{1-e^{-\tauE}} -
\dfrac{f_\epsilon}{\calH(\etaE)} \dfrac{(\partial_\eta \tausf)
  e^{-\tausf(\eta)}}{1 - e^{-\tauE}}\,.
\end{equation}
In the above equations, $\tausf(\eta) \equiv \tauE
\Heaviside{\eta-\etaE}$, $\Heaviside{x}$ being the Heaviside step
function. As discussed in Sec.~\ref{sec:angpower}, in the small angle
limit and assuming $\tauE \ll 1$ one recovers
Eq.~(\ref{eq:deltaTb}). The multipole moments read
\begin{equation}
\label{eq:Flapprox}
\begin{aligned}
F_\ell & \simeq e^{-\tauC} \Tbtilde \left[ \left( \xH \Db - \xI \DxI \right)
j_\ell(k \Delta \etaE) \right. \\ & + \left. \xH \Dv j_{\ell}''(k
\Delta \etaE) \right].
\end{aligned}
\end{equation}
Plugging this equation into Eq.~(\ref{eq:Cl21cm}) gives
Eq.~(\ref{eq:Clapprox}).

\bibliography{biblio21cm}

\end{document}